\documentclass[twocolumn]{autart} 

\usepackage{graphicx}                            
\usepackage{hyperref}
\hypersetup{hidelinks, pdfauthor=Samuel Mallick, pdfcreator=Samuel Mallick}
\usepackage{xcolor}	
\usepackage{amsfonts}   
\usepackage{amsmath}    
\usepackage{bm}        
\usepackage{algorithm}	
\usepackage{algpseudocode}
\usepackage{svg}	
\usepackage{subfig}	
\usepackage{cite}	
\usepackage{tikz}
\usepackage[utf8]{inputenc}
\usepackage{pgfplots}
\usepgfplotslibrary{groupplots,dateplot}
\usetikzlibrary{patterns,shapes.arrows}
\pgfplotsset{compat=newest}
\def\axisdefaultheight{110pt}

\DeclareMathOperator*{\argmin}{arg\,min}

\allowdisplaybreaks

\begin{document}

\begin{frontmatter}

\title{Multi-Agent Reinforcement Learning via Distributed MPC as a Function Approximator}
\newcommand{\mytitle}{Multi-Agent Reinforcement Learning via Distributed MPC as Function Approximator}
\hypersetup{hidelinks,pdfauthor=Samuel Mallick, pdfcreator=Samuel Mallick, pdftitle=\mytitle}

\thanks[corresponding]{Corresponding author S. Mallick. This paper is part of a project that has received funding from the European Research Council (ERC) under the European Union’s Horizon 2020 research and innovation programme (Grant agreement No. 101018826
- CLariNet).}

\author[]{Samuel Mallick\thanksref{corresponding}}\ead{s.h.mallick@tudelft.nl},    
\author[]{Filippo Airaldi}\ead{f.airaldi@tudelft.nl},               
\author[]{Azita Dabiri}\ead{a.dabiri@tudelft.nl},  
\author[]{Bart De Schutter}\ead{b.deschutter@tudelft.nl}

\address{Delft Center for Systems and Control, Delft University of Technology, Delft, The Netherlands}


\begin{abstract}
This paper presents a novel approach to multi-agent reinforcement learning (RL) for linear systems with convex polytopic constraints.
Existing work on RL has demonstrated the use of model predictive control (MPC) as a function approximator for the policy and value functions.
The current paper is the first work to extend this idea to the multi-agent setting.
We propose the use of a distributed MPC scheme as a function approximator, with a structure allowing for distributed learning and deployment. 
We then show that Q-learning updates can be performed distributively without introducing nonstationarity, by reconstructing a centralized learning update.
The effectiveness of the approach is demonstrated on two numerical examples. 
\end{abstract}

\end{frontmatter}

\section{Introduction}
\label{sec:introduction}
Reinforcement learning (RL) \cite{suttonReinforcementLearningIntroduction2018} has proven to be a popular approach for control of complex processes.
For large or continuous state and action spaces, function approximators are commonly used to learn representations of the policy.
Deep neural networks (DNNs) \cite{arulkumaranDeepReinforcementLearning2017} are a prevalent choice in this context; however, they often lack interpretability and are not conducive to safety verification, resulting in traditional DNN-based RL not yet widely being accepted in the control community.
Alternatively, model predictive control (MPC) is an extremely successful control paradigm \cite{borrelliPredictiveControlLinear2016}, involving solving a finite-horizon optimal control problem in a receding horizon fashion.
Extensive results on the stability and performance of MPC exist \cite{mayneConstrainedModelPredictive2000}.
However, MPC is entirely model-based, with its performance depending on an accurate system model.

The integration of MPC and RL is a promising direction for achieving safe and interpretable learning-based control \cite{hewingLearningBasedModelPredictive2020, rosoliaLearningModelPredictive2018}.
In particular, MPC has been proposed as a replacement for DNN function approximators in RL \cite{grosDataDrivenEconomicNMPC2020}.
In this context, the optimal control action and cost of the MPC optimization problem represent the policy and value function respectively.
An MPC-based policy facilitates the use of the rich theory underlying MPC for obtaining insights into the policy, and allows to deliver certificates on the resulting behavior.
\begin{color}{black}The state of the art \cite{grosDataDrivenEconomicNMPC2020, grosLearningMPCStability2022, AIRALDI20235759}, however, relies on a centralized approach with a single learning agent.
This is in general prohibitive for multi-agent systems, where centralization requires either a specific topology, with all agents connected to the central agent, or multi-hop communication across intermediate agents, where the number of hops grows with the network size. 
Additionally, centralized computation can become too complex, and requires the sharing of sensitive information, such as objective functions, with the central agent.
	
Addressing these challenges, distributed control of multi-agent systems offers computational scalability and privacy, with only neighbor-to-neighbor communication. \end{color}
Many existing works have adapted the MPC methodology to the distributed setting with distributed MPC \cite{maestreDistributedModelPredictive2014} and, likewise, RL to the multi-agent setting RL (MARL) setting, in which multiple learning agents act in a common environment.
A central challenge in MARL is that the interaction of simultaneously learning agents renders the learning target of each agent nonstationary, degrading the performance and convergence properties of single-agent algorithms \cite{busoniuComprehensiveSurveyMultiagent2008}. 
Several works have tried to circumvent this problem using a centralized training and decentralized execution paradigm \cite{alqahtaniDynamicEnergyScheduling2022, loweMultiAgentActorCriticMixed2017}.
However, centralized training is often either unrealistic or unavailable. 
Some approaches address the nonstationarity issue through communication across the network of agents during learning \cite{zhangFullyDecentralizedMultiAgent2018, suttleMultiAgentOffPolicyActorCritic2019, waiMultiAgentReinforcementLearning2018}. 
These works provide theoretical convergence guarantees, but focus on linear function approximation.
MARL has also been addressed with DNN function approximators \cite{foersterStabilisingExperienceReplay2017, guptaCooperativeMultiagentControl2017}; however, these approaches do not emphasize information exchange between agents, and suffer from the same drawbacks as DNN-based single-agent RL.
Addressing the nonstationarity of learning targets in MARL remains an open challenge.

This paper proposes the following contributions.
The use of MPC as a function approximator in RL is extended to the multi-agent setting. 
To this end we propose a structured convex distributed MPC scheme as an approximator for the policy and value functions, introducing a novel, model-based MARL approach for linear systems, free from nonstationary.
\begin{color}{black} The method is distributed in training \textit{and} deployment, with data sharing only between neighbors, irrespective of the network size and topology, thus avoiding centralized computation and multi-hop data communication.
Furthermore, privacy of sensitive information in local parameters and functions is preserved, with only state trajectories being shared, in contrast to a centralized approach where local functions are shared with the central agent.\end{color}
Thanks to the MPC-based approximation, insights into the policy can be gained from the learned components, e.g., the prediction model and constraints.
Additionally, in contrast to DNN-based approaches, it is possible to inject \textit{a priori} information, e.g., model approximations.
Furthermore, we prove a result for consensus optimization; relating the dual variables recovered distributively through the alternating direction method of multipliers (ADMM) to the optimal dual variables of the original problem, that enables the distributed learning.

The paper is structured as follows. 
Section \ref{sec:background} provides the problem description and background theory.
In Section \ref{sec:prelim_result} we present a result on the dual variables in ADMM, which will be used later on in the paper.
In Section \ref{sec:MPC_func_approx} we introduce the structured distributed MPC function approximator.
In Section \ref{sec:dist_q_learn} we propose Q-learning as the learning algorithm, and show how the parameter updates can be performed distributively.
Section \ref{sec:ex} gives illustrative examples.

\section{Preliminaries and background}\label{sec:background}
\subsection{Notation}
Define the index sets $\mathcal{M} = \{1,...,M\}$ and $\mathcal{K} = \{0,...,N-1\}$.
The symbols $t$, $k$, and $\tau$ represent time steps in an RL context, an MPC context, and iterations of an algorithm, respectively.
The symbols $s$ and $a$ refer to states and actions in an RL context, while $x$ and $u$ are used for MPC.
We use bold variables to gather variables over a prediction window, e.g., $\textbf{u} = (u^\top(0), \dots, u^\top(N-1))^\top$ and $\textbf{x} = (x^\top(0), \dots, x^\top(N))^\top$.
A vector stacking the vectors $x_i$, $i \in \mathcal{M}$, in one column is denoted $\text{col}_{i\in \mathcal{M}}(x_i)$.
The term \textit{local} describes components known only by the corresponding agent.
For simplicity we write $(x^\tau)^\top$ as $x^{\tau,\top}$.

\subsection{Problem description}\label{sec:sys_descrip}
We define a multi-agent Markov decision process (MDP) for $M$ agents as the tuple $(\mathcal{S}, \{\mathcal{A}_i\}_{i \in \mathcal{M}}, P, \{{L}_i\}_{i \in \mathcal{M}}, \mathcal{G})$. 
The set $\mathcal{S}$ is the global state set, composed of the local state sets for each agent ${\mathcal{S}} = \mathcal{S}_1 \times \dots \times \mathcal{S}_M$. 
Moreover, $\mathcal{A}_i$ is the local action set and ${L}_i$ is the local cost function for agent $i$, while $P$ describes the state transition dynamics for the whole system. 
The graph $\mathcal{G} = (\mathcal{M}, \mathcal{E})$ defines a coupling topology between agents in the network, where edges $\mathcal{E}$ are ordered pairs $(i, j)$ indicating that agent $i$ may affect the cost and state transition of agent $j$. 
Define the neighborhood of agent $i$ as $\mathcal{N}_i = \{j \in {M} | (j, i) \in \mathcal{E}, i \neq j\}$.
Note that an agent is not in its own neighborhood, i.e., $(i,i) \notin \mathcal{E}$.
We assume the graph $\mathcal{G}$ is connected.
Additionally, agents $i$ and $j$ can communicate if $i \in \mathcal{N}_j$ or $j \in \mathcal{N}_i$.

We consider agents to be linear dynamical systems with state $s_i \in \mathcal{S}_i \subseteq \mathbb{R}^n$ and control input $a_i \in \mathcal{A}_i \subseteq \mathbb{R}^m$.
The true dynamics $P$ of the network are assumed to be unknown, and we introduce an approximation of the dynamics for agent $i$, parametrized by a local parameter $\theta_i$, as
\begin{equation}
	\label{eq:nom_mod}
	\begin{aligned}
		s_i&(t+1) = f_{{\theta_i}}(s_i(t), a_i(t), \{s_j(t)\}_{j \in \mathcal{N}_i}) \\
		&= A_{i, \theta_i} s_i(t) + B_{i, \theta_i} a_i(t) + \sum_{j \in \mathcal{N}_i} A_{ij, \theta_i} s_j(t) + b_{\theta_i},
	\end{aligned}
\end{equation}
\begin{color}{black}with $b_{\theta_i} \in \mathbb{R}^n$ a constant offset allowing (1) to capture affine relationships.\end{color}

We consider a co-operative RL setting.
At time step $t$, agent $i$ observes its own state $s_{i, t} \in \mathcal{S}_i$ and takes an action with a local policy parametrized by the local parameter $\theta_i$; $\pi_{\theta_i}(s_{i, t}) = a_{i, t}$, observing the incurred local cost $L_{i, t}$ and next state $s_{i, t+1}$.
\begin{color}{black} The cooperative goal is to minimize, by modifying the parameters $\theta_i$, the discounted cost over an infinite horizon 
\begin{equation}
	\label{eq:RL_goal}
	J(\{\pi_{\theta_i}\}_{i \in \mathcal{M}}) = \mathbb{E} \left( \sum_{t=0}^{\infty} \gamma^t L_{t} \right),
\end{equation}
with $L_t = \frac{1}{M}\sum_{i \in \mathcal{M}}L_{i, t}$ the average of the agents' local costs, and $\gamma \in (0, 1]$. \end{color}
We define the global parametrization as $\theta = (\theta_1^\top,...,\theta_M^\top)^\top$.
The joint policy, parametrized by $\theta$, is then $\pi_\theta(s) = \{\pi_{\theta_i}(s_i)\}_{i \in \mathcal{M}}$.
The joint action $a = \{a_i\}_{i \in \mathcal{M}}$ is generated by the joint policy; $\pi_\theta(s) = a$, where $s$ is the joint state $s = \{s_i\}_{i \in \mathcal{M}}$.
\begin{color}{black}Note that the local policy $\pi_{\theta_i}$ is parameterized with the same parameter $\theta_i$ as the dynamics $f_{\theta_i}$ because the policy generates an action via an optimization problem in which these dynamics form equality constraints (see Section \ref{sec:MPC_func_approx}).\end{color}

\subsection{Consensus optimization}
\label{sec:ADMM_intro}
Section \ref{sec:MPC_func_approx} shows that evaluation of the proposed MPC-based policy and value functions can be posed as a consensus optimization problem and solved using the ADMM and global average consensus (GAC) algorithms.
This section provides the relevant background.

\textbf{ADMM: }
ADMM solves problems of the form
\begin{equation}
	\label{eq:ADMM_general}
	\min_{x \in \mathcal{X}, z\in \mathcal{Z}} \: \{f_\text{ADMM}(x) + g_\text{ADMM}(z) : Ax + Bz = c\}
\end{equation}
by alternating between minimization of the augmented Lagrangian, split over $x$ and $z$, and maximization of the result with respect to the multipliers $y$ as:
\begin{subequations}
	\label{eq:ADMM_general_iters}
	\begin{align}
		&x^{\tau+1} = \argmin_{x \in \mathcal{X}} \: \mathcal{L}(x, z^\tau, y^\tau) \\
		&z^{\tau+1} = \argmin_{z \in \mathcal{Z}} \: \mathcal{L}(x^{\tau+1}, z, y^\tau) \\
		&y^{\tau+1} = y^\tau + \rho(Ax^{\tau+1} + Bz^{\tau+1} - c),
	\end{align}
\end{subequations}
with $y$ the Lagrange multipliers, $\rho > 0$, and $\mathcal{L}(x, z, y) = f_\text{ADMM}(x) + g_\text{ADMM}(z) + y^\top(Ax + Bz - c) + \frac{\rho}{2}\|Ax + Bz - c\|^2_2$ the augmented Lagrangian.
We have the following convergence result:
\begin{prop}[\cite{motaProofConvergenceAlternating2018} ]
	\label{prop:ADMM}
	Assume the optimal solution set of \eqref{eq:ADMM_general} is nonempty and has optimal objective $P^\star$, the functions $f_\text{ADMM}$ and  $g_\text{ADMM}$ are convex, $\mathcal{X}$ and $\mathcal{Z}$ are convex polytopic sets, and $A$ and $B$ are full column rank.
	Then, $f_\text{ADMM}(x^\tau) + g_\text{ADMM}(z^\tau) \to P^\star$ as $\tau \to \infty$. 
	Additionally, $\{(x^\tau, z^\tau)\}_{\tau = 1}^\infty$ has a single limit point $(x^\star, z^\star)$, which solves \eqref{eq:ADMM_general}.
\end{prop}
	
Consider the following optimization problem defined over the graph $\mathcal{G}$:
\begin{subequations}\begin{align}
		\min_{x_1,\dots,x_M} &\sum_{i \in \mathcal{M}} F_i(x_i, \{x_j\}_{j \in \mathcal{N}_i})  \\
		\text{s.t.} \quad &h_i(x_i, \{x_j\}_{j \in \mathcal{N}_i}) \leq 0, \: i \in \mathcal{M} \label{eq:gen_ineq} \\
		&g_i(x_i, \{x_j\}_{j \in \mathcal{N}_i}) = 0, \: i \in \mathcal{M}, \label{eq:gen_eq}
	\end{align}\label{eq:ADMM_ex}\end{subequations}
where $F_i$, $h_i$, and $g_i$ are convex and local to agent $i$, along with its state $x_i \in \mathbb{R}^n$.
To solve this problem distributively we introduce the augmented state $\Tilde{x}_i = (x_i^\top, \text{col}^\top_{j \in \mathcal{N}_i}(x_j^{(i)}))^\top$ for agent $i$, where $x_j^{(i)}$ is a local copy of agent $j$'s state.
We then introduce a \textit{global} copy of each state $z = (z_1^\top, z_2^\top,...z_M^\top)^\top$, with $z_i$ corresponding to $x_i$.
The relevant portion of the global copies for agent $i$ is $\Tilde{z}_i = (z_i^\top, \text{col}^\top_{j \in \mathcal{N}_i}(z_j))^\top$.
Define the local feasible sets for the augmented states as
\begin{equation}
	{\mathcal{\Tilde{X}}_i} = \{\Tilde{x}_i | h_i(x_i, \{x_j^{(i)}\}_{j \in \mathcal{N}_i}) \leq 0, g_i(x_i, \{x_j^{(i)}\}_{j \in \mathcal{N}_i}) = 0\}.
\end{equation}
Problem \eqref{eq:ADMM_ex} can then be reformulated with the addition of a redundant constraint
\begin{equation}
	\label{eq:ADMM_con}
	\begin{aligned}
		\min_{\{\Tilde{x}_i \in \mathcal{\Tilde{X}}_i\}_{i \in \mathcal{M}}, z} \: &\sum_{i \in \mathcal{M}} F_i(x_i, \{x_j^{(i)}\}_{j \in \mathcal{N}_i}) \\
		\text{s.t.} \quad  &\Tilde{x}_i - \Tilde{z}_i = 0, \: i \in\mathcal{M},
	\end{aligned}
\end{equation}
which is a particular instance of \eqref{eq:ADMM_general} satisfying the assumptions in Proposition \ref{prop:ADMM} (See Appendix \ref{ap:equiv}). 
The steps in \eqref{eq:ADMM_general_iters}, when applied to \eqref{eq:ADMM_con}, reduce to \cite[Section 7.2]{boydDistributedOptimizationStatistical2010}:
\begin{subequations}
	\label{eq:ADMM_iters}
	\begin{align}
		\begin{split}
			&\Tilde{x}_i^{\tau+1} = \argmin_{\Tilde{x}_i \in \mathcal{\Tilde{X}}_i} \: F_i(x_i, \{x_j^{(i)}\}_{j \in \mathcal{N}_i}) + y_i^{\tau, \top}(\Tilde{x}_i) \\
			&\quad \quad \quad  + \frac{\rho}{2}\|\Tilde{x}_i - \Tilde{z}^\tau_i\|_2^2, \:  i \in \mathcal{M}  \label{eq:ADMM_x_update} 
		\end{split}\\
		&z_i^{\tau+1} = \frac{1}{|\mathcal{N}_i|+1} (x_i^{\tau+1} + \sum_{j \in \mathcal{N}_i}x_i^{(j), \tau+1}), \:  i \in \mathcal{M} \label{eq:ADMM_z_update}\\
		&y_i^{\tau+1} = y_i^\tau + \rho(\Tilde{x}_i^{\tau+1} - \Tilde{z}_i^{\tau+1}), \:  i \in \mathcal{M}. \label{eq:ADMM_y_update}
	\end{align}
\end{subequations}
This is a distributed procedure as each step decouples over the agents, and uses only local and neighboring information.
By Proposition \ref{prop:ADMM}, at convergence of ADMM the local variable $\Tilde{x}_i$ for each agent will contain the minimizers $x_i^\star$ of the original problem \eqref{eq:ADMM_ex}, and copies of the minimizers for neighboring agents $\{x_j^{(i),\star}\}_{j \in \mathcal{N}_i}$.

\textbf{GAC: }
The GAC algorithm allows a network of agents to agree on the average value of local variables $v_{i, 0} \in \mathbb{R}, i \in \mathcal{M}$, communicating over the graph $\mathcal{G}$.
\begin{color}{black}For each agent, the algorithm updates values as $v_i^{\tau+1} = \textbf{P}(i,i) v_i^{\tau} + \sum_{j|(i,j)\in \mathcal{E}} \textbf{P}(i,j) v_j^{\tau}$, where $\textbf{P} \in \mathbb{R}_+^{M \times M}$ is a doubly stochastic matrix, i.e., entries in each row and column sum to 1.\end{color}
The iterates converge as 
\begin{color}{black}$\lim_{\tau \to \infty} v_i^{\tau} = M^{-1} \sum_{i \in \mathcal{M}} v_{i, 0},\: i \in \mathcal{M}$, with $M$ the number of agents \cite{olfati-saberConsensusCooperationNetworked2007a}.\end{color}

\section{Local recovery of optimal dual variables from ADMM}
\label{sec:prelim_result}
In this section we provide a result linking the dual variables from the local minimization step \eqref{eq:ADMM_x_update} to a subset of the dual variables in the original problem \eqref{eq:ADMM_ex}.
This will be used later to construct the distributed learning update.
The Lagrangian of \eqref{eq:ADMM_ex} is
\begin{equation}
	\begin{aligned}
		\mathcal{L}(&\{x_i\}_{i \in \mathcal{M}}, \{\lambda_i\}_{i \in \mathcal{M}}, \{\mu_i\}_{i \in \mathcal{M}}) = \sum_{i \in \mathcal{M}} \big( f_i(x_i, \{x_j\}_{j \in \mathcal{N}_i}) \\
		&+ \lambda_i^\top h_i(x_i, \{x_j\}_{j \in \mathcal{N}_i}) + \mu_i^\top g_i(x_i, \{x_j\}_{j \in \mathcal{N}_i}) \big),
	\end{aligned}
\end{equation}
where $\lambda_i$ and $\mu_i$ are the multipliers associated with the inequality and equality constraints, for each agent.
We stress that these multipliers are related to the local constraints, and differ from the consensus multipliers $\{y_i\}_{i \in \mathcal{M}}$ used in the ADMM iterations.
Denote the optimal multipliers as $(\{\lambda^\star_i\}_{i \in \mathcal{M}}, \{\mu^\star_i\}_{i \in \mathcal{M}})$.
At iteration $\tau$ of ADMM, the Lagrangian of the local minimization step \eqref{eq:ADMM_x_update} for agent $i$ is 
\begin{equation}
	\begin{aligned}
		\mathcal{L}_i^\tau(&x_i, {\lambda}_i^\tau, {\mu}_i^\tau) = f_i(x_i, \{x_j^{(i)}\}_{j \in \mathcal{N}_i})\\
		 &\:\: + {\lambda}_i^{\tau, \top} h_i(x_i, \{x_j^{(i)}\}_{j \in \mathcal{N}_i}) + {\mu}_i^{\tau,\top} g_i(x_i, \{x_j^{(i)}\}_{j \in \mathcal{N}_i}) \\
		 &\:\:  + y_i^{\tau,\top}(\Tilde{x}_i) + \frac{\rho}{2}\|\Tilde{x}_i - \Tilde{z}^\tau_i\|_2^2.
	\end{aligned}
\end{equation}
Denote the optimal multipliers for iteration $\tau$ as $({\lambda}^{\star, \tau}_i, {\mu}^{\star, \tau}_i)$.
\begin{prop}
	\label{prop:duals}
	Assume that the assumptions in Proposition \ref{prop:ADMM} hold for problem \eqref{eq:ADMM_ex}.
	Additionally assume that the functions $F_i$ are \textit{strictly} convex.
	Then, at convergence of ADMM, the optimal multipliers from the local minimizations \eqref{eq:ADMM_x_update} converge to the corresponding subset of optimal multipliers for the original problem \eqref{eq:ADMM_ex}, i.e.,
	\begin{equation}
		({\lambda}^{\star, \tau}_i, {\mu}^{\star, \tau}_i) \to (\lambda_i^\star, \mu_i^\star), \:\:\: i \in \mathcal{M}, \:\:\: \text{as} \:\:\: \tau \to \infty.
	\end{equation}
\end{prop}
\begin{pf}
	See Appendix \ref{ap:B}.
\end{pf}

\section{Distributed MPC as a function approximator}\label{sec:MPC_func_approx}
In \cite{grosDataDrivenEconomicNMPC2020} it was shown how an MPC scheme can capture the RL policy, state-value, and action-value functions.
This section introduces a distributed counterpart, parametrized in $\theta = (\theta_1^\top,...,\theta_M^\top)^\top$, approximating these quantities for a multi-agent system.

\subsection{Parametrized distributed MPC scheme}\label{sec:dmpc_func_approx}
Consider the distributed MPC-based action-value function approximation:
\begin{subequations}
	\label{eq:DMPC_val}
	\begin{align}
		\begin{split}
				Q_\theta(s, a) &= \min_{\substack{(\{\textbf{x}_i\}, \{\textbf{u}_i\},  \{\bm{\sigma}_i\})_{i \in \mathcal{M}}}} \sum_{i \in \mathcal{M}} \Big( \beta_{\theta_i}(x_i(0))  \\
			{}&+ \sum_{k \in \mathcal{K}} \bigl( \gamma^k(l_{{\theta_i}}(x_i(k), u_i(k), \{x_j(k)\}_{j \in \mathcal{N}_i})  \\
			{}& + w^\top \sigma_i(k)) \bigr) + \gamma^N V_{\text{f},{{\theta_i}}}\big(x_i(N)\big) \Big) \label{eq:DMPC_val_cost}
		\end{split}\\
		\text{s.t.}& \quad \forall  i \in \mathcal{M}, \: \forall k \in \mathcal{K}: \nonumber \\
		&u_i(0) = a_i \label{eq:DMPC_val_IC}, \: x_i(0) = s_i \\
		x_i(k&+1) = f_{{\theta_i}}\big(x_i(k), u_i(k), \{x_j(k)\}_{j \in \mathcal{N}_i}\big) \label{eq:DMPC_val_dynam}\\
		h_{{\theta_i}}\big(&x_i(k), u_i(k), \{x_j(k)\}_{j \in \mathcal{N}_i}\big) \leq \sigma_i(k). \label{eq:DMPC_ineq}
	\end{align}
\end{subequations}
The functions $\beta_{\theta_i}$, $l_{\theta_i}$, and $V_{\text{f}, {\theta_i}}$ are the initial, stage, and terminal cost approximations respectively, $f_{\theta_i}$ is a model approximation, and $h_{\theta_i}$ is an inequality constraint function.
Each is parametrized by a local parameter $\theta_i$ and \textit{known only to the corresponding agent}, such that knowledge of the components of the MPC scheme is distributed across the agents.
Parameterized terminal constraints could be included, but are omitted here for simplicity.
For conciseness, we summarize \eqref{eq:DMPC_val} as
\begin{equation}
	\begin{aligned} \label{eq:Q_short}
		Q_\theta(s, a) = \min &\sum_{i \in \mathcal{M}} F_{\theta_i}(\textbf{x}_i, \{\textbf{x}_j\}_{j \in \mathcal{N}_i}, \textbf{u}_i, \bm{\sigma}_i) \\
		\text{s.t.}& \quad \forall  i \in \mathcal{M}, \\
		&G_{\theta_i}(s_i, a_i, \textbf{x}_i, \{\textbf{x}_j\}_{j \in \mathcal{N}_i}, \textbf{u}_i) = 0 \\
		&H_{\theta_i}(\textbf{x}_i, \{\textbf{x}_j\}_{j \in \mathcal{N}_i}, \textbf{u}_i, \bm{\sigma}_i) \leq 0.
	\end{aligned}
\end{equation}
The joint policy is then computed as
\begin{equation}
	\label{eq:policy}
	\begin{aligned}
		\pi_\theta(s) = &\argmin_{\{u(0)_i\}_{i \in \mathcal{M}}} \: \sum_{i \in \mathcal{M}} F_{\theta_i}(\textbf{x}_i, \{\textbf{x}_j\}_{j \in \mathcal{N}_i}, \textbf{u}_i, \bm{\sigma}_i)\\
		\text{s.t.}& \quad \forall  i \in \mathcal{M}, \: \forall k \in \mathcal{K}: \\ &\text{\eqref{eq:DMPC_val_dynam}--\eqref{eq:DMPC_ineq}}, \: x_i(0) = s_i
	\end{aligned}
\end{equation}
and the global value function $V_\theta(s)$ can be obtained as the optimal value of \eqref{eq:policy}, satisfying the fundamental Bellman equations \cite{grosDataDrivenEconomicNMPC2020}.

Intuitively, this structured MPC scheme approximates the local cost of each agent through the local cost functions.
Interactions between coupled agents enter the scheme via the parametrized stage costs $F_{\theta_i}$, the dynamics in $G_{\theta_i}$, and the inequality constraints in $H_{\theta_i}$.
Inexact knowledge of agents' dynamics, constraints, inter-agent interactions, and local costs can be encoded in an initial guess of the local parameters $\theta_i$.
Through learning, the cost and model representations will be modified to improve the global performance.
\begin{color}{black}The constraints in \eqref{eq:Q_short} are chosen to be affine and the functions $F_{\theta_i}$ to be strictly convex in their arguments $\textbf{x}_i, \{\textbf{x}_j\}_{j \in \mathcal{N}_i}, \textbf{u}_i,$ and $\bm{\sigma}_i$, such that \eqref{eq:Q_short} satisfies the assumptions for Propositions \ref{prop:ADMM} and \ref{prop:duals}.
This is not a strong assumption for linear systems, as the true optimal value function is convex when the stage costs are convex, as is common in the literature \cite{borrelliPredictiveControlLinear2016}.
Indeed, for quadratic stage costs, with a sufficiently long prediction horizon, a convex optimization problem can capture the true infinite horizon cost \cite{chmielewski1996constrained}.
\end{color}

\subsection{Distributed evaluation}

The ADMM and GAC algorithms can be used to solve \eqref{eq:DMPC_val} distributively.
Each agent stores a local copy of the predicted states of neighboring agents over the prediction horizon, and constructs the augmented state	$\Tilde{\textbf{x}}_i = (\textbf{x}_i^\top, \text{col}^\top_{j \in \mathcal{N}_i}(\textbf{x}_j^{(i)}))^\top$, where $\textbf{x}_j^{(i)}$ is agent $i$'s local copy of agent $j$'s predicted state over the prediction horizon.
Global copies are introduced of each state prediction $\textbf{z} = (\textbf{z}_1^\top, \textbf{z}_2^\top, \dots, \textbf{z}_M^\top)^\top$, with the relevant components of $\textbf{z}$ for agent $i$ denoted $\Tilde{\textbf{z}}_i = (\textbf{z}_i^\top, \text{col}^\top_{j \in \mathcal{N}_i}(\textbf{z}_j))^\top$.
As in Section \ref{sec:ADMM_intro}, ADMM solves \eqref{eq:Q_short} by the following iterations:
\begin{subequations}
	\begin{align}
		\begin{split}
			\Tilde{\textbf{x}}_i^{\tau+1}, \textbf{u}_i^{\tau+1}&, \bm{\sigma}_i^{\tau+1} = \argmin_{} \: F_{\theta_i}(\textbf{x}_i, \{\textbf{x}_j^{(i)}\}_{j \in \mathcal{N}_i}, \textbf{u}_i, \bm{\sigma}_i)  \\
			&\: + \sum_{k \in \mathcal{K}} y_i^{\tau,\top}(k)(\Tilde{x}_i(k))  + \frac{\rho}{2}\|\Tilde{x}_i(k) - \Tilde{z}_i(k)^\tau\|_2^2  \\
			\text{s.t.} \quad  &G_{\theta_i}(s_i, a_i, \textbf{x}_i, \{\textbf{x}_j^{(i)}\}_{j \in \mathcal{N}_i}, \textbf{u}_i) = 0 \\
			&H_{\theta_i}(\textbf{x}_i, \{\textbf{x}_j^{(i)}\}_{j \in \mathcal{N}_i}, \textbf{u}_i, \bm{\sigma}_i) \leq 0 \label{eq:ADMM_proper_x_update}
		\end{split}\\
		& \textbf{z}_i^{\tau+1} = \frac{1}{|\mathcal{N}_i|+1} (\textbf{x}_i^{\tau+1} + \sum_{j \in \mathcal{N}_i}\textbf{x}_i^{(j), \tau+1})  \label{eq:ADMM_proper_z_update}\\
		&\textbf{y}_i^{\tau+1} = \textbf{y}_i^\tau + \rho(\Tilde{\textbf{x}}_i^{\tau+1} - \Tilde{\textbf{z}}_i^{\tau+1}). \label{eq:ADMM_proper_y_update}
	\end{align}
\end{subequations}
By Proposition \ref{prop:ADMM}, as $\tau \to \infty$, the outputs of the local minimization \eqref{eq:ADMM_proper_x_update} converge to the minimizers of the original problem \eqref{eq:Q_short}, i.e., $\Tilde{\textbf{x}}_i^{\tau+1}, \textbf{u}_i^{\tau+1}, \bm{\sigma}_i^{\tau+1} \to \Tilde{\textbf{x}}_i^{\star}, \textbf{u}_i^{\star}, \bm{\sigma}_i^{\star}$.
Agents then evaluate their local objective component $F_{\theta_i}^\star = F_{\theta_i}(\textbf{x}_i^\star, \{\textbf{x}_j^{(i), \star}\}_{j \in \mathcal{N}_i}, \textbf{u}_i^\star, \bm{\sigma}_i^\star)$.
The global action-value $Q_\theta(s, a) = \sum_{i \in \mathcal{M}} F_{\theta_i}^\star$ can then be agreed upon across the network via the GAC algorithm.
Each agent makes the naive initial guess $Q_\theta(s, a) = M F_{\theta_i}^\star$, i.e., each agent assumes that all other agents have the same local cost.
The GAC algorithm gives convergence of these values, as per Section \ref{sec:ADMM_intro}, to the average $\sum_{i \in \mathcal{M}} F_{\theta_i}^\star$, which is the global action-value.

Evaluation of the global value $V_\theta(s)$ and the joint policy follows from applying the same steps to \eqref{eq:policy}.
From the optimal control sequence $\textbf{u}_i^\star$ agents evaluate their local component $a_i = \pi_{\theta_i}(s_i)= u_i^\star(0)$ of the joint policy $\pi_\theta(s)$. 
We highlight that, while the joint policy is a function of the global state $s$, the local policies are functions only of the local state $s_i$, i.e., $\pi_{\theta_i}(s_i)$, as \eqref{eq:ADMM_proper_x_update} requires knowledge only of $s_i$.

\subsection{Summary of distributed MPC as function approximator}

We briefly summarize the key points of the distributed-MPC-based approximator.
Each agent stores local knowledge of its learnable parameter $\theta_i$, and its local parametrized functions $\beta_{\theta_i}$, $l_{\theta_i}$, $V_{\text{f}, {\theta_i}}$, $f_{\theta_i}$, and $h_{\theta_i}$.
In addition, it maintains its own state prediction $\textbf{x}_i$, copies of predictions for its neighbors' states $\{\textbf{x}_j^{(i)}\}_{j \in \mathcal{N}_i}$, and an additional set of copies required for agreement. 
Finally, each agent stores the local Lagrange multipliers $\textbf{y}_i$ used in ADMM.
Evaluating  $Q_\theta$ (or $V_\theta$), from the perspective of agent $i$, is summarized in Algorithm \ref{alg:alg1}.
For simplicity, both ADMM and GAC use a fixed number of iterations in the algorithm, $T_{\text{A}}$ and $T_{\text{C}}$ respectively, with the consequences discussed in Section \ref{sec:errors}.

\begin{algorithm}
	\caption{Evaluation of $Q_\theta(s, a)$ for agent $i$.}\label{alg:alg1}
	\begin{algorithmic}[1]
		\State \textbf{Inputs}: $s_i$ and $a_i$ (not $a_i$ if evaluating \eqref{eq:policy}).
		\State \textbf{Initialize}: $\{\textbf{x}_j^{(i)}\}_{j \in \mathcal{N}_i} \gets 0$, $\textbf{z}_i \gets 0$, $\textbf{y}_i \gets 0$.
		\For{$\tau = 0, 1, \dots,  T_{\text{A}}$}
		\State Get $\Tilde{\textbf{x}}_i^{\tau+1}, \textbf{u}_i^{\tau+1}, \bm{\sigma}_i^{\tau+1}$ via \eqref{eq:ADMM_proper_x_update}.
		\State For $j \in \mathcal{N}_i$, send $\textbf{x}_j^{(i), \tau+1}$ and receive $\textbf{x}_i^{(j), \tau+1}$.
		\State Perform averaging step \eqref{eq:ADMM_proper_z_update}.
		\State For $j \in \mathcal{N}_i$, send $\textbf{z}_i^{\tau+1}$ and receive $\textbf{z}_j^{\tau+1}$.
		\State Perform local multipliers update \eqref{eq:ADMM_proper_y_update}.
		\EndFor
		\State  $\Bar{Q}_\theta(s, a) \gets M F_{\theta_i}({\textbf{x}}_i^{T_{\text{A}}}, \{\textbf{x}_j^{(i), T_\text{A}}\}_{j \in \mathcal{N}_i}, \textbf{u}_i^{T_{\text{A}}}, \bm{\sigma}_i^{T_{\text{A}}})$
		\State Perform $T_\text{C}$ iterations GAC, with initial guess $\Bar{Q}_\theta(s, a)$, to agree on $Q_\theta(s, a)$. \label{alg:cons_step}
		\State \textbf{Outputs}: Global state-value $Q_\theta(s, a)$ $\big($local action $a_i \gets u_i^{T_{\text{A}}}(0)$ if evaluating \eqref{eq:policy}$\big)$.
	\end{algorithmic}
\end{algorithm}

\section{Distributed Q-learning}\label{sec:dist_q_learn}
In this section we show that using Q-learning as the RL algorithm to learn the local parameters $\theta_i$ enables a distributed learning update that avoids nonstationarity.
Q-learning \cite{suttonReinforcementLearningIntroduction2018} adjusts, at each time step $t$, the global parameters $\theta = (\theta_1^\top,...,\theta_M^\top)^\top$ as
\begin{equation}
	\label{eq:q_learn_cent}
	\begin{aligned}
		&\delta_t = L_{t} + \gamma V_\theta(s_{t+1}) - Q_\theta(s_t, a_t) \\
		&\theta \gets \theta + \alpha \delta_t \nabla_\theta Q_\theta(s_t, a_t),
	\end{aligned}
\end{equation}
where $\alpha \in \mathbb{R}$ is a learning rate and $\delta_t \in \mathbb{R}$ is the temporal-difference (TD) error at time step $t$.

\subsection{Separable Q-learning updates}
The update \eqref{eq:q_learn_cent} appears to be centralized due to the global components $a_t$, $L_{t}$, $s_t$, $s_{t+1}$, and $\theta$.
However, the structure of \eqref{eq:DMPC_val} allows decomposition into local updates of $\theta_i$ for each agent.
Addressing first the TD error $\delta_t$, we have shown in Section \ref{sec:MPC_func_approx} that $V_\theta(s)$ and $Q_\theta(s, a)$ can be evaluated distributively. 
The globally averaged cost $L_{t}$ is the average of all agents' locally incurred costs $L_{i, t}$, and can be shared across the network using the GAC algorithm.
This can be incorporated into line \ref{alg:cons_step} of Algorithm \ref{alg:alg1}, with no additional communication overhead.

We now address the gradient term $\nabla_\theta Q_\theta(s_t, a_t)$.
The Lagrangian of \eqref{eq:DMPC_val} is 
\begin{equation}
	\label{eq:lagrangian}
	\begin{aligned}
		\mathcal{L}_\theta&(s, a, p) = \sum_{i \in \mathcal{M}} \bigg( F_{\theta_i}(\textbf{x}_i, \{\textbf{x}_j\}_{j \in \mathcal{N}_i}, \textbf{u}_i, \bm{\sigma}_i) \\
		&+ \bm{\lambda}_i^\top G_{\theta_i}(s_i, a_i, \textbf{x}_i, \{\textbf{x}_j\}_{j \in \mathcal{N}_i}, \textbf{u}_i) \\
		&+ \bm{\mu}_i^\top H_{\theta_i}(\textbf{x}_i, \{\textbf{x}_j\}_{j \in \mathcal{N}_i}, \textbf{u}_i, \bm{\sigma}_i) \bigg),
	\end{aligned}
\end{equation}
where $\bm{\lambda}_i$ and $\bm{\mu}_i$ are the multiplier vectors associated with the equality and inequality constraints, respectively, and the primal and dual variables are grouped as
\begin{equation}
	\begin{aligned}
		p = (\{\textbf{x}_i\}, \{\textbf{u}_i\}, \{\bm{\sigma}_i\}, \{\bm{\lambda}_i\}, \{\bm{\mu}_i\})_{i \in \mathcal{M}}.
	\end{aligned}
\end{equation}
We stress again that the multipliers in $p$ are associated with the constraints of \eqref{eq:DMPC_val}, and are unrelated to the multipliers $\textbf{y}_i$ used in the ADMM procedure.
Via sensitivity analysis \cite{buskensSensitivityAnalysisRealTime2001}, the gradient $\nabla_\theta Q_\theta(s, a)$ coincides with the gradient of the Lagrangian at the optimal primal and dual variables $p^\star$:
\begin{equation}
	\label{eq:Q_sens}
	\nabla_\theta Q_\theta(s, a) = \frac{\partial \mathcal{L}_\theta(s, a, p^\star)}{\partial \theta}.
\end{equation}
Observe that \eqref{eq:lagrangian} is separable over parameters $\theta_i$, states $s_i$, actions $a_i$, and subsets of the primal and dual variables $p_i$, i.e., $\mathcal{L}_\theta(s, a, p) = \sum_{i \in \mathcal{M}} \mathcal{L}_{\theta_i}(s_i, a_i, p_i)$,
where $\mathcal{L}_{\theta_i}$ is the $i$-th term within the sum \eqref{eq:lagrangian}, and $p_i = (\textbf{x}_i, \textbf{u}_i, \bm{\sigma}_i, \{\textbf{x}_j\}_{j \in \mathcal{N}_i}, \bm{\lambda}_i, \bm{\mu}_i)$.
We then express \eqref{eq:Q_sens} as
\begin{equation}
	\frac{\partial \sum_{i \in \mathcal{M}} \mathcal{L}_{\theta_i}(s_i, a_i, p_i^\star)}{\partial \theta} = \begin{bmatrix}
		\frac{\partial \mathcal{L}_{\theta_1}(s_1, a_1, p_1^\star)}{\partial \theta_1} \\
		\vdots\\
		\frac{\partial \mathcal{L}_{\theta_M}(s_M, a_M, p_M^\star)}{\partial \theta_M}
	\end{bmatrix},
\end{equation}
and the centralized parameter update \eqref{eq:q_learn_cent} can be hence unpacked into $M$ local updates:
\begin{equation}
	\label{eq:Q_learn_local}
	\theta_i \gets \theta_i + \alpha \delta_t \frac{\partial \mathcal{L}_{\theta_i}(s_{i, t}, a_{i, t}, p_i^\star)}{\partial \theta_i},\: i \in \mathcal{M}.
\end{equation}

What then remains to be shown is that the subset of optimal primal and dual variables $p_i^\star$ for \eqref{eq:DMPC_val} is available locally to agent $i$.
By Proposition \ref{prop:ADMM}, at convergence of ADMM, the local minimization \eqref{eq:ADMM_proper_x_update} returns the minimizers $(\textbf{x}_i^\star, \textbf{u}_i^\star, \bm{\sigma}_i^\star)$ and local copies of the minimizers of neighboring agents $\{\textbf{x}_j^{(i), \star}\}_{j \in \mathcal{N}_i} = \{\textbf{x}_j^{\star}\}_{j \in \mathcal{N}_i}$.
Each agent hence has local knowledge of the optimal \textit{primal} values in $p_i^\star$.
For the optimal \textit{dual} values, by Proposition \ref{prop:duals}, at convergence of ADMM, the optimal dual variables of the local minimization \eqref{eq:ADMM_proper_x_update} for agent $i$ are equal to the relevant subset of optimal dual variables for the original problem, i.e., $(\bm{\lambda}^\star_i, \bm{\mu}^\star_i)$.
Each agent hence has local knowledge of the optimal \textit{dual} values in $p_i^\star$ as well.
Agents can therefore perform the local update \eqref{eq:Q_learn_local}.
Nonstationarity is avoided as the local updates reconstruct the centralized update of the whole network \eqref{eq:q_learn_cent}.

\subsection{Implementation details}\label{sec:errors}
In this section we discuss some auxiliary details in the implementation of our proposed approach.
\begin{itemize}
\item\begin{color}{black} Exploration can be added to the approach in, e.g., an epsilon-greedy fashion, in which case the agent adds a random perturbation to the objective \eqref{eq:DMPC_val_cost} with some probability $\epsilon_i$, where $\epsilon_i$ decreases as training progresses \cite{AIRALDI20235759,zanonSafeReinforcementLearning2021}.
	This causes a perturbation in the joint policy.
	It is assumed that the system, with exploration injected, is sufficiently persistently exciting.
\end{color}
		
\item The finite termination of ADMM and the GAC algorithm introduces errors in the evaluation of $V_\theta$ and $Q_\theta$, $\pi_\theta$, and $p^\star$, and could lead to instability in the learning.
To counter this, large numbers of ADMM and GAC iterations may be used in a simulated learning phase, when there are no constraints on computation time between actions.
This is desirable as primal and dual variables with high precision are needed to reliably compute the sensitivity \eqref{eq:Q_sens}.
On the contrary, at deployment, sensitivities are not required and only the policy must be evaluated.
Thus, the iteration numbers can be reduced, as ADMM often converges to modest accuracy within few iterations \cite{boydDistributedOptimizationStatistical2010}.
Additionally, experience replay (ER) \cite{AIRALDI20235759, lin1992self} can improve learning stability by using an average of past observations when calculating the gradient and the TD error.
In our method ER requires no extra mechanism as agents can maintain a local history of values for $\delta$ and $\nabla_{\theta_i} \mathcal{L}_{\theta_i}(s, p_i^\star)$.
In our numerical experiments, we found the learning to succeed with ER and modest numbers of ADMM and GAC iterations.

\item \begin{color}{black}
The variables $\alpha > 0$ and $\gamma \in (0, 1]$ are hyperparameters; as the distributed update fully reconstructs the centralized update, existing methods for selecting these parameters in centralized learning apply directly, e.g., see \cite[Chapter 9.6]{suttonReinforcementLearningIntroduction2018}.\end{color}
\end{itemize}

\section{Example}\label{sec:ex}
\begin{color}{black}This section presents two numerical examples. 
Source code and simulation results can be found at \url{https://github.com/SamuelMallick/dmpcrl-concept}.\end{color}
\subsection{Academic example}
We modify the system from \cite{grosDataDrivenEconomicNMPC2020}, forming a three-agent system with state coupling in a chain, i.e., $1 \leftrightarrow 2 \leftrightarrow 3$, \begin{color}{black}with real (unknown) dynamics $s_i(t+1) = A_i s_i(t) + B_i a_i(t) + \sum_{j \in \mathcal{N}_i} A_{ij} s_j(t) + [e_i(t), 0]^\top$ where
\begin{equation} \label{eq:true_mod}
	A_i = \begin{bmatrix}
		0.9 & 0.35 \\ 0 & 1.1
	\end{bmatrix}, \{A_{ij}\}_{j \in \mathcal{N}_i} = \begin{bmatrix}
	0 & 0 \\ 0 & -0.1
	\end{bmatrix},
\end{equation}
with $B = [0.0813, 0.2]^\top$ and $e_i(t)$ uniformly distributed over the interval $[-0.1, 0]$.
\begin{color}{black} The RL task is to drive the states and control towards the origin while avoiding violations of the state constraints $\underline{s} \leq s_i \leq \overline{s}$, with local costs as:
	\begin{equation}
		\begin{aligned}
			&L_{i, t}(s_i, a_i) = \|s_i\|_2^2 + \frac{1}{2}\|u_i\|_2^2 \\
			 &{} \: + \max\big((0, \omega^\top (\underline{s} - s_i)\big) + \max\big((0, \omega^\top (s_i - \overline{s})\big),
		\end{aligned} 
	\end{equation}
with $\omega = [10^2, 10^2]^\top$, $\underline{s} = [0, -1]^\top$, and $\overline{s} = [1, 1]^\top$.
Non-positive noise on the first state biases that term towards violating the lower bound of zero.
This renders the task challenging, as the agents must regulate the state to zero, and yet driving the first dimension to zero will result in constraint violations due to the noise.\end{color}
The true model is unknown, with agents instead knowing a uniform distribution of models, containing the true model:
\begin{equation}
	\label{eq:inac_mod}
	\begin{aligned}
		&\Hat{A}_i = \begin{bmatrix}
			1 + a_{1,i} & 0.25 + a_{2,i}\\ 0 & 1+ a_{3,i}
		\end{bmatrix}, \{\Hat{A}_{ij}\}_{j \in \mathcal{N}_i} = \begin{bmatrix}
			0 & 0 \\ 0 & c_i
		\end{bmatrix},
	\end{aligned}
\end{equation}
with $\Hat{B}_i = [0.0312 + b_{1, i}, 0.25 + b_{2, i}]^\top$ and, for all $i$, the random variables distributed uniformly as $a_{1,i}, a_{2,i}, a_{3,i} \in [-0.1, 0.1]$, $b_{1,i} \in [0, 0.075]$, $b_{2,i} \in [-0.075, 0]$ and $c_i \in [-0.1, 0]$.\end{color}
We implement the following distributed MPC scheme:
\begin{equation*}
\begin{aligned}\label{eq:academic_dmpc}
		\min_{\substack{\{(\textbf{x}_i, \textbf{u}_i, \bm{\sigma}_i)\}_{i \in \mathcal{M}}}} &\sum_{i \in \mathcal{M}}\Big( V_{i, 0} + \sum_{k \in \mathcal{K}} f_i^\top \begin{bmatrix}
			x_i(k) \\ u_i(k)
		\end{bmatrix} \\
		&\hspace{-0.5cm}+ \frac{1}{2}\gamma^k \big( \|x_i(k)\|^2 + \frac{1}{2}\|u_i(k)\|^2 + \omega^\top \sigma_i(k) \big) \Big)\\
		\text{s.t.}& \quad \forall  i \in \mathcal{M}, \: \forall k \in \mathcal{K}: \nonumber \\
		&x_i(k+1) = A_i x_i(k) + B_i u_i(k) \\
		&\quad\quad+ \sum_{j \in \mathcal{N}_i} A_{ij} x_j(k) + b_i\\
		&\underline{s} + \underline{x}_i - \sigma_i(k) \leq x_i(k) \leq \overline{s} + \overline{x}_i + \sigma_i(k)\\
	&-1 \leq u_i(k) \leq 1, \: x_i(0) = s_i
\end{aligned}
\end{equation*}
where $M = 3$, $\mathcal{K} = \{0,\dots,9\}$, and $\gamma = 0.9$.
The learnable parameters for each agents are then $\theta_i = (V_{i,0}, \underline{x}_i, \overline{x}_i, b_i, f_i, A_i, B_i, \{A_{ij}\}_{j \in \mathcal{N}_i})$.
The initial values for $A_i, B_i, \{A_{ij}\}_{j \in \mathcal{N}_i}$ are the inaccurate model \eqref{eq:inac_mod} with the random variables set to zero. All other learnable parameters are initialized to zero.

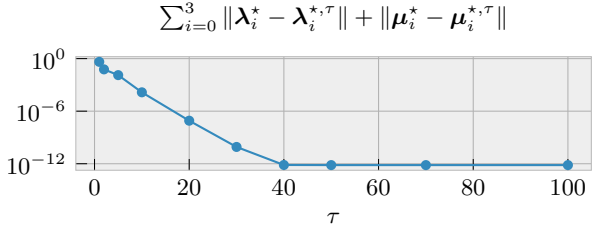
\begin{figure}
		\centering
\begin{tikzpicture}

\definecolor{darkgray178}{RGB}{178,178,178}
\definecolor{silver188}{RGB}{188,188,188}
\definecolor{steelblue52138189}{RGB}{52,138,189}
\definecolor{whitesmoke238}{RGB}{238,238,238}

\begin{semilogyaxis}[
font=\small,
height=0.8*\axisdefaultheight,
width=\axisdefaultwidth,
axis background/.style={fill=whitesmoke238},
axis line style={silver188},
tick pos=left,
tick scale binop=\times,
x grid style={darkgray178},
xlabel={\(\displaystyle \tau\)},
xmajorgrids,
xmin=-3.95, xmax=104.95,
xtick style={color=black},
y grid style={darkgray178},
title={$\sum_{i=0}^3 \| \bm{\lambda}_i^\star - \bm{\lambda}_i^{\star, \tau} \| + \| \bm{\mu}_i^\star - \bm{\mu}_i^{\star, \tau}\|$},
ymajorgrids,
ymin=1.79495967785803e-13, ymax=1.66688414588261,
ytick style={color=black},
]
\addplot [thick, steelblue52138189, mark=*, mark size=1.5, mark options={solid}]
table {%
1 0.429000359361865
2 0.059175278170437
5 0.0130359840105544
10 0.000141966589749687
20 8.17799909442446e-08
30 8.2414368400772e-11
40 7.2052758791538e-13
50 6.97746117107097e-13
70 6.97432942473674e-13
100 6.97940731663806e-13
};
\end{semilogyaxis}

\end{tikzpicture}
		\vspace{-0.3cm}
		\caption{Accuracy of the dual variables recovered by Proposition \ref{prop:duals} as a function of the ADMM iteration index $\tau$.}
		\label{fig:dual_errors}
\end{figure}

To illustrate Proposition \ref{prop:duals}, for a given global state $s$, Figure \ref{fig:dual_errors} shows the error between the true optimal dual variables and those recovered from evaluating the MPC scheme with ADMM, as a function of the ADMM iteration index $\tau$.
The locally recovered dual variables are close to the true values, with the error initially decreasing with the iteration index.

\begin{figure}
	\centering
	\includesvg[width=1\columnwidth]{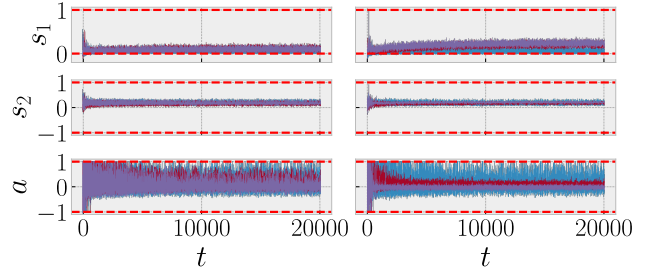}
	\vspace{-0.5cm}
	\caption{Centralized (left) and distributed (right). Evolution of the states and inputs during training. Agent 1 (blue), agent 2 (red), agent 3 (purple), and bounds (dashed).}
	\label{fig:num_states}
\end{figure}

\begin{figure}
	\centering
	\includesvg[width=1\columnwidth]{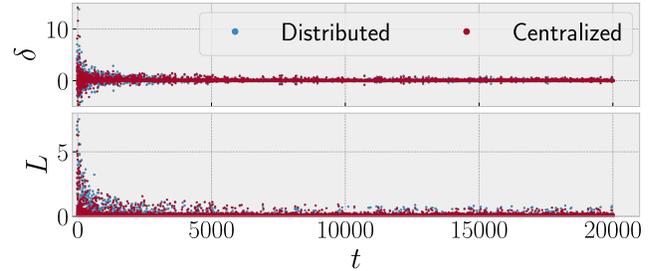}
	\vspace{-0.75cm}
	\caption{Evolution of TD errors (top) and stage costs (bottom) during training.}
	\label{fig:num_TD}
\end{figure}

\begin{figure}
	\centering
	\includesvg[width=1\columnwidth]{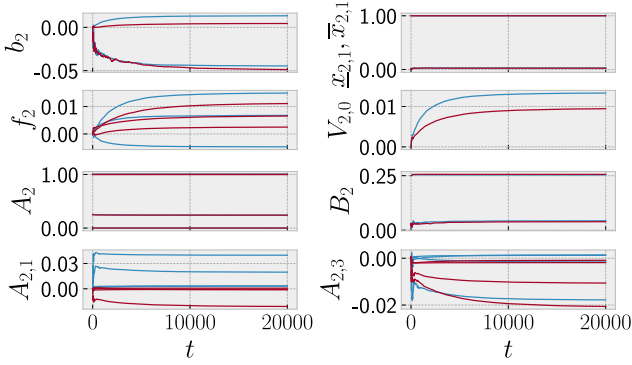}
	\vspace{-0.75cm}
	\caption{Evolution of learnable parameters for agent $2$ during training. Distributed (blue) and centralized (red).}
	\label{fig:num_pars}
\end{figure}

We now compare the learning performance of our distributed approach with the centralized approach inspired from \cite{grosDataDrivenEconomicNMPC2020}, using the same MPC scheme, and the same learning hyperparameters for both.
We use an exploration probability of $\epsilon_i = 0.7$, exponentially decaying with rate 0.99, with local costs perturbed uniformly over the interval $[-1, 1]$.
We use a learning rate of $\alpha = 6e^{-5}$, exponentially decaying with rate 0.9996, and ER with an average over 15 past samples at an update rate of every 2 time steps.
For the distributed approach we use 50 iterations in ADMM and 100 iterations in GAC.
These values were tuned to keep the iterations low without introducing significant approximation errors.
Figure \ref{fig:num_states} shows the state and input trajectories for the three agents during training.
Figure \ref{fig:num_TD} shows the evolution of the global TD error and the collective cost.
Figure \ref{fig:num_pars} shows the learnable parameters of the second agent during training (similar convergence profiles are observed for the other agents).
It is seen that the behavior of the distributed approach is similar to that of the centralized approach, reducing the TD error and costs incurred, with the centralized approach converging slightly faster.
The costs are reduced by maintaining the first state of each agent above zero to prevent expensive violations of the state constraint due to coupling and noise.

\begin{color}{black} We compare the performance of the distributed policy against both a distributed nominal MPC controller (NMPC) and a distributed stochastic MPC controller (SMPC) based on the scenario approach \cite{schildbachScenarioApproachStochastic2014}.
The nominal MPC controller uses the inexact model \eqref{eq:inac_mod}, with all random variables set to 0.
For the SMPC controller we consider the case where the true model is unknown and $N=25$ samples of the model distribution \eqref{eq:inac_mod} and the noise distribution are used to account for the uncertainty. 
The number of samples was manually tuned to balance conservativeness and robustness \cite{schildbachScenarioApproachStochastic2014}.
We also consider the case where the exact model \eqref{eq:true_mod} is known, and only the noise distribution is sampled.
The controllers are compared in Figure 5, showing the closed-loop cost accumulated over 100 time steps.
The learned policy can be seen to improve from a performance initially comparable to NMPC, learning to outperform SMPC, and approaching the performance of SMPC with a perfect model.
We highlight that our approach retains the computational complexity of NMPC, while SMPC is significantly more complex, optimizing over $N$ copies of the state trajectories. 
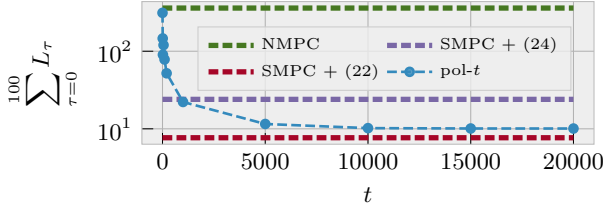
\begin{figure}
	\centering
\begin{tikzpicture}

\definecolor{darkgray178}{RGB}{178,178,178}
\definecolor{darkolivegreen7012033}{RGB}{70,120,33}
\definecolor{firebrick166640}{RGB}{166,6,40}
\definecolor{lightgray204}{RGB}{204,204,204}
\definecolor{silver188}{RGB}{188,188,188}
\definecolor{slategray122104166}{RGB}{122,104,166}
\definecolor{steelblue52138189}{RGB}{52,138,189}
\definecolor{whitesmoke238}{RGB}{238,238,238}

\newcommand{\lw}{2}

\begin{axis}[
font=\small,
height=0.9*\axisdefaultheight,
width=0.9*\axisdefaultwidth,
axis background/.style={fill=whitesmoke238},
axis line style={silver188},
legend cell align={left},
legend style={
  fill opacity=0.8,
  draw opacity=1,
  text opacity=1,
  at={(0.95,0.6)},
  anchor=east,
  draw=lightgray204,
  fill=whitesmoke238,
  legend columns=2,
  font = \tiny
},
log basis y={10},
tick pos=left,
x grid style={darkgray178},
xlabel={\(\displaystyle t\)},
xmajorgrids,
xmin=-1000, xmax=21000,
xtick style={color=black},
tick label style={/pgf/number format/fixed},
scaled ticks=false,
xtick={-5000,0,5000,10000,15000,20000,25000},
xticklabels={
  \(\displaystyle {\ensuremath{-}5000}\),
  \(\displaystyle {0}\),
  \(\displaystyle {5000}\),
  \(\displaystyle {10000}\),
  \(\displaystyle {15000}\),
  \(\displaystyle {20000}\),
  \(\displaystyle {25000}\)
},
y grid style={darkgray178},
ylabel={\(\displaystyle \sum_{\tau = 0}^{100} L_\tau\)},
ymajorgrids,
ymin=6.33142583834917, ymax=436.614095299082,
ymode=log,
ytick style={color=black},
ytick={0.1,1,10,100,1000,10000},
yticklabels={
  \(\displaystyle {10^{-1}}\),
  \(\displaystyle {10^{0}}\),
  \(\displaystyle {10^{1}}\),
  \(\displaystyle {10^{2}}\),
  \(\displaystyle {10^{3}}\),
  \(\displaystyle {10^{4}}\)
}
]
\addplot [darkolivegreen7012033, dash pattern=on 3.7pt off 1.6pt, line width=\lw]
table {%
0 360.184646407951
20000 360.184646407951
};
\addlegendentry{NMPC}
\addplot [slategray122104166, dash pattern=on 3.7pt off 1.6pt, line width=\lw]
table {%
0 23.973033151068
20000 23.973033151068
};
\addlegendentry{SMPC + \eqref{eq:inac_mod}}
\addplot [firebrick166640, dash pattern=on 3.7pt off 1.6pt, line width=\lw]
table {%
0 7.67492393674399
20000 7.67492393674399
};
\addlegendentry{SMPC + \eqref{eq:true_mod}}
\addplot [steelblue52138189, dash pattern=on 3.7pt off 1.6pt, mark=*, mark size=1.5, mark options={solid}, line width=1]
table {%
0 312.397390247417
10 146.250353765369
20 90.2305150011611
50 120.12132011415
100 78.3391023459313
200 51.9750600935795
1000 22.1814389941827
5000 11.5417506069651
10000 10.2072615675652
15000 10.0969785698873
20000 10.0856950287377
};
\addlegendentry{pol-$t$}
\end{axis}

\end{tikzpicture}
	\vspace{-0.4cm}
	\caption{NMPC and SMPC compared against policy at training time steps $t$ (log scale).}
	\label{fig:pol_eval}
	\end{figure}
\end{color}

\subsection{Power systems example}
Automatic generation control is a popular case study for distributed control \cite{feleCoalitionalControlSelfOrganizing2018}.
We consider the benchmark system presented, along with test scenarios, in \cite{riversoHycon2BenchmarkPower2012}.
The network is a four-area power system whose continuous-time linearized dynamics for sub-system $i$ are $\Dot{x}_i = A_i \big(\{{P}_{ij}\}_{j \in \mathcal{N}_i} \big)x_i + B_i u_i + L_i \Delta P_{L_i} + \sum_{j \in \mathcal{N}_i} A_{ij}\big(\{{P}_{ij}\}_{j \in \mathcal{N}_i} \big) x_j$,
where $x_i = (\Delta \phi_i, \Delta \omega_i, \Delta P_{m_i}, \Delta P_{v_i})^\top$ contains the rotor's angular displacement $\Delta \phi_i$, the rotating mass' speed $\Delta \omega_i$, the mechanical power $\Delta P_{m_i}$, and the steam valve's position $\Delta P_{v_i}$.
Moreover, $\Delta P_{L_i}$ is the local power load, the control action $u_i$ is local power generation, and $\mathcal{N}_i$ is the set of neighboring sub-systems, i.e., those connected directly through tie-lines.
The agents are again coupled as a chain.
The $P_{ij}$ term, which enters the matrices linearly, is a synchronizing coefficient for the tie-line between sub-systems $i$ and $j$.
For the sake of space, we refer to \cite{riversoHycon2BenchmarkPower2012} for the full definitions of the dynamics matrices.
The local continuous dynamics are discretized, treating the coupling and loads as exogenous inputs, with the discrete-time matrices denoted $\Hat{A}_{i}$, $\Hat{B}_i$, $\Hat{A}_{ij}$, and $\Hat{L}_i$.
In our experiments we use the discrete-time model in the MPC controllers, and the continuous-time model for the underlying system.

We consider a scenario where the step changes $(+0.15, -0.15, +0.12, -0.24, 0.28)$ occur in the local loads $(\Delta P_{L_1},\Delta P_{L_2},\Delta P_{L_3},\Delta P_{L_3},\Delta P_{L_4})$ at times $(5, 15, 20, 40, 40)$, with the initial load values all zero.
As in \cite{feleCoalitionalControlSelfOrganizing2018}, we take the case where local power production is insufficient to balance the local loads, requiring power transfer between areas.
The control constraints are then $(\Bar{u}_1, \Bar{u}_2, \Bar{u}_3,\Bar{u}_4) = (0.2, 0.1, 0.3, 0.1)$, where $|u_i| \leq \Bar{u}_i$.
Additionally the angular displacements are constrained as $|\Delta \phi_i| \leq 0.1$.
Each sub-system has a learning agent.
Agents know the nominal value of $\Delta P_{L_i}$; however, the real loads are subject to uniform additive noise in the range $[-0.1, 0.1]$.
Exact knowledge of $P_{ij}$ is unknown, with agents having initial approximations of the true values $\Hat{P}_{ij} = P_{ij} + d_{ij}$, where $d_{ij} \sim \text{U}(-2, 2)$.
For the RL task, we take the performance criteria from \cite{riversoHycon2BenchmarkPower2012}, regulating the states and control input to the time-varying equilibria $\bar{x}_i = (0, 0, \Delta P_{L_i}, \Delta P_{L_i})^\top$ and $\bar{u}_i = \Delta P_{L_i}$, while minimizing power transfer between sub-systems and avoiding expensive constraint violations.
We implement the following distributed MPC scheme:
\begin{align*}
	\begin{split}
		\min_{\{(\textbf{x}_i, \textbf{u}_i, \bm{\sigma}_i)\}_{i \in \mathcal{M}}} &\sum_{i \in \mathcal{M}} V_{i, 0} + \sum_{k \in \mathcal{K}} f_i^\top \begin{bmatrix}
			x_i(k) \\ u_i(k)
		\end{bmatrix} \\
		&+ \gamma^k((x_i(k)^\top - \bar{x}^\top_i) Q_{x_i} (x_i(k) - \bar{x}_i) \\
		&+ (u_i(k)^\top - \bar{u}_i^\top) Q_{u_i} (u_i(k)- \bar{u}_i) \\
		&+ \omega^\top \sigma_i(k)) 
	\end{split}\\
	\text{s.t.} \quad &x_i(0) = s_i \quad i \in \mathcal{M} \\
	\begin{split}
		&x_i(k+1) = \Hat{A}_i \big(\{\Hat{P}_{ij}\}_{j \in \mathcal{N}_i} \big) x_i(k) + \Hat{B}_i u_i(k) \\
		&\quad\:+  \hat{L}_i \Delta P_{L_i}  + \sum_{j \in \mathcal{N}_i} \Hat{A}_{ij} \big(\{\Hat{P}_{ij}\}_{j \in \mathcal{N}_i} \big) x_j(k) \\
		&\quad\:+ 0.1 b_i
	\end{split}\\
	\begin{split}
		&-0.1+ \underline{\Delta \phi}_i - \sigma_i(k) \leq \Delta \phi_i(k) \\
		&\quad\: \leq 0.1  + \overline{\Delta \phi}_i + \sigma_i(k)
	\end{split}\\
	&-\Bar{u}_i \leq u_i(k) \leq \Bar{u}_i \quad k \in \mathcal{K}\quad i \in \mathcal{M},
\end{align*}
where $M = 4$, $\mathcal{K} = \{0,\dots,4\}$, $\omega = (50^2, 50^2)^\top$, and $\gamma = 0.9$.
The learnable parameters for each agents are $\theta_i = (V_{i,0}, \underline{\Delta \phi}_i, \overline{\Delta \phi}_i, b_i, f_i, Q_{x_i}, Q_{u_i}, \{\Hat{P}_{ij}\}_{j \in \mathcal{N}_i})$.
The $\Hat{P}_{ij}$ values are initialized with the initial approximated values known by the agents, while $Q_{x_i}$ and $Q_{u_i}$ are initialized with the values given in the MPC controller in \cite{riversoHycon2BenchmarkPower2012}.
All other parameters are initialized to zero.
The agents are trained episodically.
We use an exploration probability of $\epsilon = 0.5$, exponentially decaying with rate 0.8, with local costs perturbed uniformly over the interval $[-0.04, 0.04]$.
We use a learning rate of $\alpha = 7e^{-7}$, exponentially decaying with rate 0.98, and experience replay over 3 episodes at an update rate of once per episode.
Again, we use the tuned number of iterations: 50 iterations in ADMM and 100 iterations in GAC.

\begin{figure}
	\centering
	\input{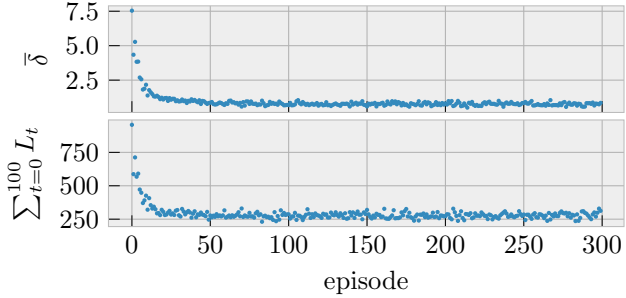}
	\vspace{-20pt}
	\caption{Average TD error (top) and the return (bottom) per episode during training.}
	\label{fig:power_TD}
\end{figure}

Figure \ref{fig:power_TD} shows the average TD error $\Bar{\delta}$ and return $\sum_{t=0}^{100} L_t$ per episode.
The distributed learning succeeds in reducing the TD error and improving the overall performance of the network.
Figure \ref{fig:power_states} shows $\Delta \phi_4$ and the deviation in power flow from area 3 to area 4, i.e., $\Delta P_{\text{tie}, 3, 4} = P_{34}(\Delta \phi_3 - \Delta \phi_4)$, for the first and last episodes of training.
The local load for area 4 is not satisfiable by local generation, requiring power flow from area 3, and driving $\Delta \phi_4$ towards the bound.
Initially, due to noise and an incorrect model, this leads to constraint violations.
After training, it can be seen that agent 3 learned to provide a larger power surge to agent 4, attenuating the decrease in $\Delta \phi_4$, and avoiding constraint violations.
We also compare the performance of the resulting policy against a distributed nominal MPC controller and a distributed stochastic MPC controller, based on the Scenario approach \cite{schildbachScenarioApproachStochastic2014}, to account for the noise on the load.
Both use the true model and stage costs tuned as in \cite{riversoHycon2BenchmarkPower2012}.
Figure \ref{fig:power_eval} demonstrates the performance and number of constraint violations over 100 episodes of the policy, the stochastic MPC controller, and the nominal MPC controller.
Stochastic MPC improves on the performance of nominal MPC, as it better avoids violations.
The proposed approach, despite starting from an incorrect model of the dynamics, distributively learned a policy that achieves zero violations and improves the performance over the stochastic and nominal MPC controllers.
\begin{figure}
	\centering
	\input{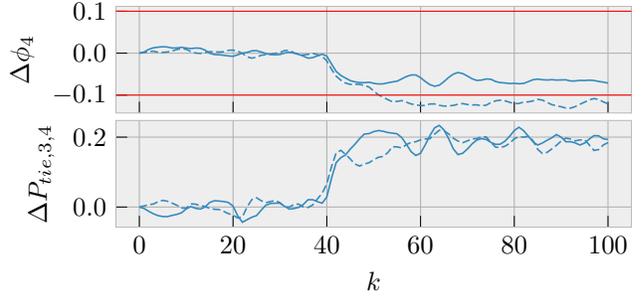}
	\vspace{-20pt}
	\caption{$\Delta \phi_4$ (top) and $\Delta P_{\text{tie},3,4}$ (bottom) for the first episode (dashed) and the last episode (solid) of training.}
	\label{fig:power_states}
\end{figure}
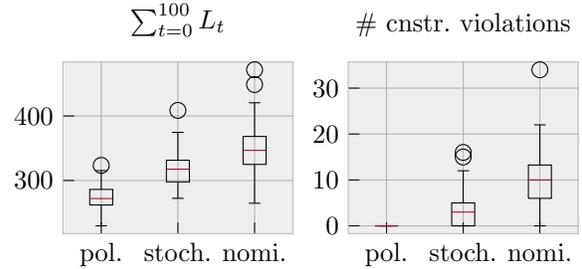
\begin{figure}
	\centering 
	\subfloat{
\begin{tikzpicture}

\definecolor{darkgray178}{RGB}{178,178,178}
\definecolor{firebrick166640}{RGB}{166,6,40}
\definecolor{silver188}{RGB}{188,188,188}
\definecolor{whitesmoke238}{RGB}{238,238,238}

\begin{axis}[
width=0.55\columnwidth,
height=\axisdefaultheight,
axis background/.style={fill=whitesmoke238},
axis line style={silver188},
tick pos=left,
tick scale binop=\times,
x grid style={darkgray178},
xmajorgrids,
xmin=-0.5, xmax=2.5,
xtick style={color=black},
xtick={0,1,2},
xticklabels={pol.,stoch., nomi.},
y grid style={darkgray178},
title={\(\sum_{t=0}^{100} L_t\)},
ymajorgrids,
ymin=217.549131985686, ymax=483.783748322577,
ytick style={color=black},
ytick={200,300,400,500},
yticklabels={
  \(\displaystyle {200}\),
  \(\displaystyle {300}\),
  \(\displaystyle {400}\),
  \(\displaystyle {500}\)
}
]
\addplot [black]
table {%
-0.15 262.008685000063
0.15 262.008685000063
0.15 286.15181990184
-0.15 286.15181990184
-0.15 262.008685000063
};
\addplot [black]
table {%
0 262.008685000063
0 229.650705455545
};
\addplot [black]
table {%
0 286.15181990184
0 315.362925575074
};
\addplot [black]
table {%
-0.075 229.650705455545
0.075 229.650705455545
};
\addplot [black]
table {%
-0.075 315.362925575074
0.075 315.362925575074
};
\addplot [black, mark=o, mark size=3, mark options={solid,fill opacity=0}, only marks]
table {%
0 323.402506679597
};
\addplot [black]
table {%
0.85 297.837930472522
1.15 297.837930472522
1.15 331.403013860209
0.85 331.403013860209
0.85 297.837930472522
};
\addplot [black]
table {%
1 297.837930472522
1 272.350874046006
};
\addplot [black]
table {%
1 331.403013860209
1 374.637125571572
};
\addplot [black]
table {%
0.925 272.350874046006
1.075 272.350874046006
};
\addplot [black]
table {%
0.925 374.637125571572
1.075 374.637125571572
};
\addplot [black, mark=o, mark size=3, mark options={solid,fill opacity=0}, only marks]
table {%
1 408.67198246025
};
\addplot [black]
table {%
1.85 324.955989856157
2.15 324.955989856157
2.15 368.295975783019
1.85 368.295975783019
1.85 324.955989856157
};
\addplot [black]
table {%
2 324.955989856157
2 264.715649461118
};
\addplot [black]
table {%
2 368.295975783019
2 420.464339377745
};
\addplot [black]
table {%
1.925 264.715649461118
2.075 264.715649461118
};
\addplot [black]
table {%
1.925 420.464339377745
2.075 420.464339377745
};
\addplot [black, mark=o, mark size=3, mark options={solid,fill opacity=0}, only marks]
table {%
2 448.77028481635
2 471.682174852718
};
\addplot [firebrick166640]
table {%
-0.15 271.97041702211
0.15 271.97041702211
};
\addplot [firebrick166640]
table {%
0.85 317.552627780442
1.15 317.552627780442
};
\addplot [firebrick166640]
table {%
1.85 346.699509876666
2.15 346.699509876666
};
\end{axis}

\end{tikzpicture}}
	\subfloat{
\begin{tikzpicture}

\definecolor{darkgray178}{RGB}{178,178,178}
\definecolor{firebrick166640}{RGB}{166,6,40}
\definecolor{silver188}{RGB}{188,188,188}
\definecolor{whitesmoke238}{RGB}{238,238,238}

\begin{axis}[
width=0.55\columnwidth,
height=\axisdefaultheight,
axis background/.style={fill=whitesmoke238},
axis line style={silver188},
tick pos=left,
tick scale binop=\times,
x grid style={darkgray178},
xmajorgrids,
xmin=0.5, xmax=3.5,
xtick style={color=black},
xtick={1,2,3},
xticklabels={pol.,stoch.,nomi.},
y grid style={darkgray178},
title={\# cnstr. violations},
ymajorgrids,
ymin=-1.7, ymax=35.7,
ytick style={color=black},
ytick={-10,0,10,20,30,40},
yticklabels={
  \(\displaystyle {\ensuremath{-}10}\),
  \(\displaystyle {0}\),
  \(\displaystyle {10}\),
  \(\displaystyle {20}\),
  \(\displaystyle {30}\),
  \(\displaystyle {40}\)
}
]
\addplot [black]
table {%
0.85 0
1.15 0
1.15 0
0.85 0
0.85 0
};
\addplot [black]
table {%
1 0
1 0
};
\addplot [black]
table {%
1 0
1 0
};
\addplot [black]
table {%
0.925 0
1.075 0
};
\addplot [black]
table {%
0.925 0
1.075 0
};
\addplot [black]
table {%
1.85 0
2.15 0
2.15 5
1.85 5
1.85 0
};
\addplot [black]
table {%
2 0
2 0
};
\addplot [black]
table {%
2 5
2 12
};
\addplot [black]
table {%
1.925 0
2.075 0
};
\addplot [black]
table {%
1.925 12
2.075 12
};
\addplot [black, mark=o, mark size=3, mark options={solid,fill opacity=0}, only marks]
table {%
2 15
2 16
};
\addplot [black]
table {%
2.85 6
3.15 6
3.15 13.25
2.85 13.25
2.85 6
};
\addplot [black]
table {%
3 6
3 0
};
\addplot [black]
table {%
3 13.25
3 22
};
\addplot [black]
table {%
2.925 0
3.075 0
};
\addplot [black]
table {%
2.925 22
3.075 22
};
\addplot [black, mark=o, mark size=3, mark options={solid,fill opacity=0}, only marks]
table {%
3 34
};
\addplot [firebrick166640]
table {%
0.85 0
1.15 0
};
\addplot [firebrick166640]
table {%
1.85 3
2.15 3
};
\addplot [firebrick166640]
table {%
2.85 10
3.15 10
};
\end{axis}

\end{tikzpicture}}
	\vspace{-10pt}
	\caption{Box-plots of performance and the number of constraint violations per episode, over 100 episodes, of the learned policy, stochastic MPC, and nominal MPC.}
	\label{fig:power_eval}
\end{figure}

\section{Conclusions}
This paper has extended the idea of using MPC-based RL to the multi-agent setting.
We have proposed a novel approach to MARL via the use of distributed MPC as a distributed function approximator within Q-learning.
A result on the optimal dual variables in ADMM is first presented.
The structure of the distributed MPC function approximator is then detailed, and is shown to enable a distributed evaluation of the global value functions and the local policies. 
Finally, by using Q-learning to update the parametrization of the MPC scheme, the parameter updates can also be performed fully distributively.
The effectiveness of the newly proposed method is demonstrated on a numerical example.

\begin{color}{black} A limitation of the proposed approach is the convexity requirement on the MPC scheme, restricting the class of systems for whom the optimal policy can be captured to linear systems with convex cost functions.
Additionally, as both the ADMM and GAC algorithms use iterative communication between neighbors, the approach is less suited for applications with high communication costs.\end{color}
Future work will look at extending the idea to other RL methods, such as policy-based approaches, and a formal analysis on the propagation of errors from the finite termination of the ADMM and GAC algorithms.
\appendix
\section*{Appendix}
	\section{Equivalence of \eqref{eq:ADMM_general} and \eqref{eq:ADMM_con}} \label{ap:equiv}
	Define the aggregate of the local augmented states as $\Bar{x} = \text{col}_{i \in \mathcal{M}}(\Tilde{x}_i)$ and the feasible set as $\Bar{\mathcal{X}} = \Tilde{\mathcal{X}}_1 \times \dots \times \Tilde{\mathcal{X}}_M$.
	Problem \eqref{eq:ADMM_con} can then be written in the form of \eqref{eq:ADMM_general} as
	\begin{equation}
		\label{eq:equiv}
		\min_{\Bar{x} \in \Bar{\mathcal{X}}, z} \{f_\text{ADMM}(\Bar{x}) + g_\text{ADMM}(z): \: A\Bar{x} + Bz = c\},
	\end{equation}
	where $f_\text{ADMM}(\Bar{x}) = \sum_{i \in \mathcal{M}} F_i(\Tilde{x}_i)$, $g_\text{ADMM}(\cdot)$ is the zero map, and $c = \mathbf{0}$. Furthermore, $A = \mathbf{I}_{q \times q}$, $q = \sum_{i \in \mathcal{M}} n(|\mathcal{N}_i| + 1)$, and $B \in \mathbb{R}^{q \times nM}$ is a block matrix that contains negative identity matrix entries picking out the corresponding global states, with all other entries being zero matrices. 
	Clearly, matrix $A$ is full column rank.
	For $B$, we note that every column has at least one identity entry, and that by rearranging its rows we can express it as $[-\mathbf{I}_{nM \times nM}, B_+^\top]^\top$, where $B_+$ contains the remaining rows. 
	Therefore, $B$ is full column rank.
	The assumptions in Proposition \ref{prop:ADMM} are hence satisfied.
	
\section{Proof of Proposition \ref{prop:duals}}
\label{ap:B}
	The proof involves showing that the Karush-Kuhn-Tucker (KKT) conditions for the local minimization problems \eqref{eq:ADMM_x_update}, when aggregated for all agents, are equivalent to the KKT conditions of the original problem \eqref{eq:ADMM_ex}.
	Then, due to strict convexity of both \eqref{eq:ADMM_x_update} and \eqref{eq:ADMM_ex}, the primal and dual solutions are unique, and the equivalence between the optimal dual variables follows.
	We write the optimal primal and dual variables of \eqref{eq:ADMM_ex} as $(\{x_i^\star\}_{i \in \mathcal{M}}, \{\lambda^\star_i\}_{i \in \mathcal{M}}, \{\mu^\star_i\}_{i \in \mathcal{M}})$, and the optimal primal and dual variables of \eqref{eq:ADMM_x_update} for agent $i$, \textit{at convergence of ADMM}, as $(x_i^\star, \{x_j^{(i), \star}\}_{j \in \mathcal{N}_i}, \hat{\lambda}_i^\star, \hat{\mu}_i^\star)$.
	Primal variable equivalence is given by Proposition \ref{prop:ADMM}, i.e., the $x_i^\star$'s are equivalent for both problems.
	Also, optimal local copies are equal to their true values 
	\begin{equation}
		\label{eq:copy_equiv}
		\{x_j^{(i), \star}\}_{j \in \mathcal{N}_i} = \{{x}_j^{\star}\}_{j \in \mathcal{N}_i} \quad i \in \mathcal{M}.
	\end{equation}
	
	First, we state the KKT conditions of \eqref{eq:ADMM_ex}.
	These are
	\begin{subequations}
		\label{eq:KKT}
		\begin{align}
			\begin{split}
				&h_i(x_i^\star, \{x_j^\star\}_{j \in \mathcal{N}_i}) \leq 0 \quad i \in \mathcal{M} \\
				&g_i(x_i^\star, \{x_j^\star\}_{j \in \mathcal{N}_i}) = 0 \quad i \in \mathcal{M} \label{eq:KKT_prim}
			\end{split}\\
			&\lambda_i^\star \geq 0 \quad i \in \mathcal{M} \label{eq:KKT_dual}\\
			&\lambda_i^{\star, \top} h_i(x_i^\star, \{x_j^\star\}_{j \in \mathcal{N}_i}) = 0 \quad i \in \mathcal{M} \label{eq:KKT_comp}\\
			&\nabla_x \Big( \sum_{i \in \mathcal{M}}  F_i(x_i, \{x_j\}_{j \in \mathcal{N}_i})  + \lambda_i^{\top} h_i(x_i, \{x_j\}_{j \in \mathcal{N}_i}) \nonumber \\
			&+ \mu_i^{\top} g_i(x_i, \{x_j\}_{j \in \mathcal{N}_i}) \Big)|_{\substack{\{({x}_i, \lambda_i, \mu_i)\}_{i \in \mathcal{M}} \\
					= \{({x}_i^\star, \lambda_i^\star, \mu_i^\star)\}_{i \in \mathcal{M}}}} = 0 \label{eq:KKT_stat},
		\end{align}
	\end{subequations}
	where $x = (x_1^\top,\dots,x_M^\top)^\top$.
	Now consider the composition of the KKT conditions of each agent's local minimization \eqref{eq:ADMM_x_update}.
	First, primal feasibility, dual feasibility, and complementary slackness for agent $i$:
	\begin{equation}
		\label{eq:local_KKT_primal}
		\begin{aligned}
			&h_i(x_i^\star, \{x_j^{(i),\star}\}_{j \in \mathcal{N}_i}) \leq 0, \quad g_i(x_i^\star, \{x_j^{(i),\star}\}\}_{j \in \mathcal{N}_i}) = 0, \\
			&\hat{\lambda}_i^\star \geq 0, \quad \hat{\lambda}_i^{\star, \top} h_i(x_i^\star, \{x_j^{(i),\star}\}_{j \in \mathcal{N}_i}) = 0.
		\end{aligned}
	\end{equation}
	Inserting the equivalence \eqref{eq:copy_equiv} into \eqref{eq:local_KKT_primal}, we find that the composition of these conditions for all agents are the original problem's conditions \eqref{eq:KKT_prim} - \eqref{eq:KKT_comp}.
	
	For ease of exposition in showing the equivalence of the stationarity conditions, let us define
	\begin{equation}
		\begin{aligned}
			\mathcal{F}_i&(x_i, \{x_j\}_{j \in \mathcal{N}_i}, \lambda_i, \mu_i) = F_i(x_i, \{x_j\}_{j \in \mathcal{N}_i}) \\
			&+ \lambda_i^{\top} h_i(x_i, \{x_j\}_{j \in \mathcal{N}_i}) + \mu_i^{\top} g_i(x_i, \{x_j\}_{j \in \mathcal{N}_i}).
		\end{aligned}
	\end{equation}
	Moreover, we group the primal and dual variables as $p = (\{x_i\}_{i \in \mathcal{M}}, \{\lambda_i\}_{i \in \mathcal{M}}, \{\mu_i\}_{i \in \mathcal{M}})$, such that the stationarity condition \eqref{eq:KKT_stat} reads
	\begin{equation}
		\label{eq:stat}
		\nabla_x \Big( \sum_{i \in \mathcal{M}} \mathcal{F}_i(x_i, \{x_j\}_{j \in \mathcal{N}_i}, \lambda_i, \mu_i)\Big)\Big|_{p = p^\star} = 0.
	\end{equation}
	By decomposing the derivative operator and considering each row $l$ of \eqref{eq:stat} we have
	\begin{equation}
		\label{eq:row_l}
		\bigg( \frac{\partial}{\partial x_l} \sum_{i \in \mathcal{M}} \mathcal{F}_i(x_i, \{x_j\}_{j \in \mathcal{N}_i}, \lambda_i, \mu_i)\bigg)\bigg|_{p = p^\star} = 0.
	\end{equation}
	Observing that $\frac{\partial}{\partial x_l} \mathcal{F}_i(x_i, \{x_j\}_{j \in \mathcal{N}_i}, \lambda_i, \mu_i) = 0$ if $i \notin \mathcal{N}_l \cup l$, from \eqref{eq:row_l} we have
	\begin{equation}
		\begin{aligned}
			\label{eq:row_proof}
			\bigg( \frac{\partial}{\partial x_l} &\mathcal{F}_l(x_l, \{x_j\}_{j \in \mathcal{N}_l}, \lambda_l, \mu_l) \bigg)\bigg|_{p = p^\star} \\
			&+ \sum_{i \in \mathcal{N}_l} \bigg( \frac{\partial}{\partial x_l} \mathcal{F}_i(x_i, \{x_j\}_{j \in \mathcal{N}_i}, \lambda_i, \mu_i) \bigg)\bigg|_{p = p^\star} = 0.
		\end{aligned}
	\end{equation}
	We highlight that the partial derivative terms within the sum are not zero, as $x_l$ is one of the elements of $\{x_j\}_{j \in \mathcal{N}_i}$.
	We will now show that the condition \eqref{eq:row_proof} for each $l$ can be reconstructed by a unique subset of the KKT conditions arising in the agents' local minimizations. 
	Consider the stationarity conditions of the agents' local minimizations.
	The primal and dual variables for agent $i$ are grouped as $p_i = (\Tilde{x}_i, \hat{\lambda}_i, \hat{\mu}_i)$.
	The stationarity condition for agent $i$'s local minimization can be written as
	\begin{equation}
		\begin{aligned}
			\nabla_{\Tilde{x}_i} \big( &\mathcal{F}_i(x_i, \{x_j^{(i)}\}_{j \in \mathcal{N}_i}, \hat{\lambda}_i, \hat{\mu}_i) \\
			&+y_i^{\tau, \top} \Tilde{x}_i + \frac{\rho}{2}\|\Tilde{x}_i - \Tilde{z}_i^\tau\|_2^2\big)\big|_{p_i = p_i^\star}  = 0.
		\end{aligned}
	\end{equation}
	Evaluating the derivative of the second term, we have
	\begin{equation}
		\begin{aligned}
			\nabla_{\Tilde{x}_i} \big( &\mathcal{F}_i(x_i, \{x_j^{(i)}\}_{j \in \mathcal{N}_i}, \hat{\lambda}_i, \hat{\mu}_i)\big)\big|_{p_i = p_i^\star} \\
			&+ y_i^{\tau} + \rho (\Tilde{x}_i^\star - \Tilde{z}_i^\tau)  = 0.
		\end{aligned}
	\end{equation}
	From Proposition \ref{prop:ADMM}, we have that at convergence of ADMM, $\Tilde{x}_i^\star - \Tilde{z}_i^\star = 0$.
	Decomposing the derivative operator we have
	\begin{equation}
		\label{eq:local_rows}
		\begin{aligned}
			\Bigg(
			\begin{bmatrix}
				\frac{\partial}{\partial x_i} \\ \text{col}_{j \in \mathcal{N}_i} (\frac{\partial}{\partial x_j^{(i)}})
			\end{bmatrix}  &\mathcal{F}_i(x_i, \{x_j^{(i)}\}_{j \in \mathcal{N}_i}, \hat{\lambda}_i, \hat{\mu}_i) \Bigg)\Bigg|_{p_i = p_i^\star} \\
			&+ \begin{bmatrix}
				y_i^{\tau} \\ \text{col}_{j \in \mathcal{N}_i}(y_j^{(i), \tau}) 
			\end{bmatrix} = 0
		\end{aligned}
	\end{equation}
	We denote this as $\Psi_i = \begin{bmatrix}
	\Psi_i^{(i), \top} &
	\text{col}_{j \in \mathcal{N}_i}^\top (\Psi_j^{(i)})
	\end{bmatrix}^\top$, where $\Psi_j^{(i)}$ is the row of \eqref{eq:local_rows} corresponding to the partial derivative with respect to $x_j^{(i)}$.
	Let us sum all rows of the agents' local stationarity conditions which correspond to the partial derivative with respect to $x_l$, or a copy of $x_l$, to get
	\begin{equation}\label{eq:row_sum}
		\Psi_{l}^{(l)} + \sum_{i \in \mathcal{N}_l} \Psi_i^{(l)} = 0.
	\end{equation}
	The result is equal to zero as each term in the sum is zero.
	The sum expression reads
	\begin{equation}
		\begin{aligned}
			&\bigg( \frac{\partial}{\partial x_l} \mathcal{F}_l(x_l, \{x_j^{(l)}\}_{j \in \mathcal{N}_l}, \hat{\lambda}_l, \hat{\mu}_l) \bigg) \bigg|_{p_l = p_l^\star} \\
			&+ \sum_{i \in \mathcal{N}_l} \bigg( \frac{\partial}{\partial x_l^{(i)}} \mathcal{F}_i(x_i, \{x_j^{(i)}\}_{j \in \mathcal{N}_i}, \hat{\lambda}_i, \hat{\mu}_i) \bigg)\bigg|_{p_i = p_i^*}  \\
			&+ y_l^{\tau} + \sum_{i \in \mathcal{N}_l} y_l^{(i), \tau} = 0,
		\end{aligned}
	\end{equation}
	where again we highlight that the partial derivative terms in the sum are not zero, as $l \in \mathcal{N}_i$.
	It is shown in \cite{boydDistributedOptimizationStatistical2010} that the sum of the dual ADMM variables corresponding to the same global variables is zero after the first iteration of the procedure, i.e., $y_l^{\tau} + \sum_{i \in \mathcal{N}_l} y_l^{(i), \tau} = 0$, $\tau > 1$.
	Hence, at convergence, we have
	\begin{equation}
		\begin{aligned}
			&\bigg( \frac{\partial}{\partial x_l} \mathcal{F}_l(x_l, \{x_j^{(l)}\}_{j \in \mathcal{N}_l}, \hat{\lambda}_l, \hat{\mu}_l) \bigg) \bigg|_{p_l = p_l^\star} \\
			&+ \sum_{i \in \mathcal{N}_l} \bigg( \frac{\partial}{\partial x_l^{(i)}} \mathcal{F}_i(x_i, \{x_j^{(i)}\}_{j \in \mathcal{N}_i}, \hat{\lambda}_i, \hat{\mu}_i) \bigg)\bigg|_{p_i = p_i^\star} = 0.
		\end{aligned}
	\end{equation}
	Substituting the equivalence of the local copies \eqref{eq:copy_equiv} and observing that
	\begin{equation}
		\begin{aligned}
			\bigg( \frac{\partial}{\partial x_l^{(i)}} &\mathcal{F}_i(x_i, \{x_j^{(i)}\}_{j \in \mathcal{N}_i}, \hat{\lambda}_i, \hat{\mu}_i) \bigg)\bigg|_{p_l = p_l^\star} = \\
			&\bigg( \frac{\partial}{\partial x_l} \mathcal{F}_i(x_i, \{x_j\}_{j \in \mathcal{N}_i},  \hat{\lambda}_i, \hat{\mu}_i) \bigg)\bigg|_{p_l = p_l^\star},
		\end{aligned}
	\end{equation}
	we obtain the equivalent stationarity condition from the $l$-th row of \eqref{eq:stat}.
	Repeating the sum in \eqref{eq:row_sum} for each agent, we reconstruct the equivalent rows of \eqref{eq:stat}. 
	We then have equivalence between the stationarity conditions on $(\{\hat{\lambda}_i^\star\}_{i \in \mathcal{M}}, \{\hat{\mu}_i^\star\}_{i \in \mathcal{M}})$ and on $(\{\lambda^\star_i\}_{i \in \mathcal{M}}, \{\mu^\star_i\}_{i \in \mathcal{M}})$.
	This concludes the proof.
	

\bibliographystyle{plain}   
\bibliography{references} 

\begin{thebibliography}{10}

\bibitem{AIRALDI20235759}
F.~Airaldi, B.~De~Schutter, and Dabiri A.
\newblock Learning safety in model-based reinforcement learning using {MPC} and
  gaussian processes.
\newblock {\em IFAC-PapersOnLine}, 56(2):5759--5764, 2023.
\newblock 22nd IFAC World Congress.

\bibitem{alqahtaniDynamicEnergyScheduling2022}
M.~Alqahtani, M.~J. Scott, and M.~Hu.
\newblock Dynamic energy scheduling and routing of a large fleet of electric
  vehicles using multi-agent reinforcement learning.
\newblock {\em Computers \& Industrial Engineering}, 169:108180, 2022.

\bibitem{arulkumaranDeepReinforcementLearning2017}
K.~Arulkumaran, M.~P. Deisenroth, M.~Brundage, and A.~N. Bharath.
\newblock Deep reinforcement learning: A brief survey.
\newblock {\em IEEE Signal Processing Magazine}, 34(6):26--38, 2017.

\bibitem{borrelliPredictiveControlLinear2016}
F.~Borrelli, A.~Bemporad, and M.~Morari.
\newblock {\em Predictive Control for Linear and Hybrid Systems}.
\newblock Cambridge University Press, 2016.

\bibitem{boydDistributedOptimizationStatistical2010}
S.~Boyd.
\newblock Distributed optimization and statistical learning via the alternating
  direction method of multipliers.
\newblock {\em Foundations and Trends in Machine Learning}, 3(1):1--122, 2010.

\bibitem{buskensSensitivityAnalysisRealTime2001}
C.~B{\"u}skens and H.~Maurer.
\newblock Sensitivity analysis and real-time optimization of parametric
  nonlinear programming problems.
\newblock In {\em Online Optimization of Large Scale Systems}, pages 3--16.
  Springer, Berlin, Heidelberg, 2001.

\bibitem{busoniuComprehensiveSurveyMultiagent2008}
L.~Busoniu, R.~Babuska, and B.~De~Schutter.
\newblock A comprehensive survey of multiagent reinforcement learning.
\newblock {\em IEEE Transactions on Systems, Man, and Cybernetics, Part C
  (Applications and Reviews)}, 38(2):156--172, 2008.

\bibitem{chmielewski1996constrained}
Donald Chmielewski and V~Manousiouthakis.
\newblock On constrained infinite-time linear quadratic optimal control.
\newblock {\em Systems \& Control Letters}, 29(3):121--129, 1996.

\bibitem{feleCoalitionalControlSelfOrganizing2018}
F.~Fele, E.~Debada, J.~M. Maestre, and E.~F. Camacho.
\newblock Coalitional control for self-organizing agents.
\newblock {\em IEEE Transactions on Automatic Control}, 63(9):2883--2897, 2018.

\bibitem{foersterStabilisingExperienceReplay2017}
J.~Foerster, N.~Nardelli, G.~Farquhar, T.~Afouras, P.~H.~S. Torr, P.~Kohli, and
  S.~Whiteson.
\newblock Stabilising experience replay for deep multi-agent reinforcement
  learning.
\newblock In {\em International Conference on Machine Learning. PMLR}, pages
  1146--1155, 2017.

\bibitem{grosDataDrivenEconomicNMPC2020}
S.~Gros and M.~Zanon.
\newblock Data-driven economic {NMPC} using reinforcement learning.
\newblock {\em IEEE Transactions on Automatic Control}, 65(2):636--648, 2020.

\bibitem{grosLearningMPCStability2022}
S.~Gros and M.~Zanon.
\newblock Learning for {MPC} with stability \& safety guarantees.
\newblock {\em Automatica}, 146:110598, 2022.

\bibitem{guptaCooperativeMultiagentControl2017}
J.~K. Gupta, M.~Egorov, and M.~Kochenderfer.
\newblock Cooperative multi-agent control using deep reinforcement learning.
\newblock {\em Autonomous Agents and Multiagent Systems}, pages 66--83, 2017.

\bibitem{hewingLearningBasedModelPredictive2020}
L.~Hewing, K.~P. Wabersich, M.~Menner, and M.~N. Zeilinger.
\newblock Learning-based model predictive control: Toward safe learning in
  control.
\newblock {\em Annual Review of Control, Robotics, and Autonomous Systems},
  3(1):269--296, 2020.

\bibitem{lin1992self}
Long-Ji Lin.
\newblock Self-improving reactive agents based on reinforcement learning,
  planning and teaching.
\newblock {\em Machine learning}, 8:293--321, 1992.

\bibitem{loweMultiAgentActorCriticMixed2017}
R.~Lowe, Y.~Wu, A.~Tamar, J.~Harb, P.~Abbeel, and I.~Mordatch.
\newblock Multi-agent actor-critic for mixed cooperative-competitive
  environments.
\newblock {\em Advances in Neural Information Processing Systems}, 30, 2017.

\bibitem{maestreDistributedModelPredictive2014}
J.~M. Maestre and R.~R. Negenborn, editors.
\newblock {\em Distributed Model Predictive Control Made Easy}.
\newblock Dordrecht, The Netherlands: Springer, 2014.

\bibitem{mayneConstrainedModelPredictive2000}
D.~Q. Mayne, J.~B. Rawlings, C.~V. Rao, and P.~O.~M. Scokaert.
\newblock Constrained model predictive control: Stability and optimality.
\newblock {\em Automatica}, 36(6):789--814, 2000.

\bibitem{motaProofConvergenceAlternating2018}
J.~F.~C. Mota, João M.~F. Xavier, P.~M.~Q. Aguiar, and M.~Püschel.
\newblock A proof of convergence for the alternating direction method of
  multipliers applied to polyhedral-constrained functions.
\newblock {\em arXiv:1112.2295}, 2018.

\bibitem{olfati-saberConsensusCooperationNetworked2007a}
R.~Olfati-Saber, J.~A. Fax, and R.~M. Murray.
\newblock Consensus and cooperation in networked multi-agent systems.
\newblock {\em Proceedings of the IEEE}, 95(1):215--233, 2007.

\bibitem{riversoHycon2BenchmarkPower2012}
S.~Riverso and G.~{Ferrari-Trecate}.
\newblock Hycon2 benchmark: Power network system.
\newblock {\em arXiv:1207.2000}, 2012.

\bibitem{rosoliaLearningModelPredictive2018}
U.~Rosolia and F.~Borrelli.
\newblock Learning model predictive control for iterative tasks. {A}
  data-driven control framework.
\newblock {\em IEEE Transactions on Automatic Control}, 63(7):1883--1896, 2018.

\bibitem{schildbachScenarioApproachStochastic2014}
G.~Schildbach, L.~Fagiano, C.~Frei, and M.~Morari.
\newblock The scenario approach for stochastic model predictive control with
  bounds on closed-loop constraint violations.
\newblock {\em Automatica}, 50(12):3009--3018, 2014.

\bibitem{suttleMultiAgentOffPolicyActorCritic2019}
W.~Suttle, Z.~Yang, K.~Zhang, Z.~Wang, T.~Basar, and J.~Liu.
\newblock A multi-agent off-policy actor-critic algorithm for distributed
  reinforcement learning.
\newblock {\em arXiv:1903.06372}, 2019.

\bibitem{suttonReinforcementLearningIntroduction2018}
R.~S. Sutton and A.~G. Barto.
\newblock {\em Reinforcement Learning: An Introduction}.
\newblock MIT Press, 2018.

\bibitem{waiMultiAgentReinforcementLearning2018}
H.~Wai, Z.~Yang, Z.~Wang, and M.~Hong.
\newblock Multi-agent reinforcement learning via double averaging primal-dual
  optimization.
\newblock {\em Advances in Neural Information Processing Systems}, 31, 2018.

\bibitem{zanonSafeReinforcementLearning2021}
M.~Zanon and S.~Gros.
\newblock Safe reinforcement learning using robust {MPC}.
\newblock {\em IEEE Transactions on Automatic Control}, 66(8):3638--3652, 2021.

\bibitem{zhangFullyDecentralizedMultiAgent2018}
K.~Zhang, Z.~Yang, H.~Liu, T.~Zhang, and T.~Basar.
\newblock Fully decentralized multi-agent reinforcement learning with networked
  agents.
\newblock In {\em International Conference on Machine Learning. PMLR}, pages
  5872--5881, 2018.

\end{thebibliography}
\end{document}